RESEARCH ARTICLE

# User interface for in-vehicle systems with on-wheel finger spreading gestures and head-up displays

Sang Hun Lee * and Se-One Yoon

Graduate School of Automotive Engineering, Kookmin University, Seoul 02707, Republic of Korea

*Corresponding author. E-mail: shlee@kookmin.ac.kr  http://orcid.org/0000-0001-8888-2201

**Abstract**

Interacting with an in-vehicle system through a central console is known to induce visual and biomechanical distractions, thereby delaying the danger recognition and response times of the driver and significantly increasing the risk of an accident. To address this problem, various hand gestures have been developed. Although such gestures can reduce visual demand, they are limited in number, lack passive feedback, and can be vague and imprecise, difficult to understand and remember, and culture-bound. To overcome these limitations, we developed a novel on-wheel finger spreading gestural interface combined with a head-up display (HUD) allowing the user to choose a menu displayed in the HUD with a gesture. This interface displays audio and air conditioning functions on the central console of a HUD and enables their control using a specific number of fingers while keeping both hands on the steering wheel. We compared the effectiveness of the newly proposed hybrid interface against a traditional tactile interface for a central console using objective measurements and subjective evaluations regarding both the vehicle and driver behaviour. A total of 32 subjects were recruited to conduct experiments on a driving simulator equipped with the proposed interface under various scenarios. The results showed that the proposed interface was approximately 20% faster in emergency response than the traditional interface, whereas its performance in maintaining vehicle speed and lane was not significantly different from that of the traditional one.

*Keywords*: user interface; in-vehicle system; hand gesture; head-up display; human–vehicle interaction; driver distraction

## 1. Introduction

Driving is a complex task, usually requiring the complete attention resources of the driver, and thus performing other activities simultaneously may lead to a major decline in driving performance. The National Highway Traffic Safety Administration (NHTSA) classifies driver distractions into four types: visual, biomechanical, auditory, and cognitive (Ranney, Garrott, & Goodman, 2001). Visual distractions occur when a driver is looking elsewhere and therefore not paying full attention to the road. Biomechanical distractions occur when the driver uses their body for tasks other than driving, such as drinking, smoking, or interacting with in-vehicle systems. Auditory distractions occur when the driver uses devices such as smartphones, listens to the radio, or talks to a passenger. Finally, cognitive distractions occur when the driver is thinking about not only the road but also other things that might affect their driving (e.g. children running into the road chasing after a ball).

According to the NHTSA traffic accident database, 25–30% of all traffic accidents and 78% of collision accidents that occur in the United States are caused by driver distraction (Stutts, Reinfurt, Staplin, & Rodgman, 2001; Neale, Dingus, Klauer, Sudweeks, & Goodman, 2005). Actual vehicle driver data show that drivers taking their eyes off the road for more than 2 s increase the probability of an accident, whereas performing complex tasks triples the risk of a collision (Llaneras, 2000). Approximately 50% of accidents from driver distractions are caused by the use of smartphones and in-vehicle infotainment systems, such as navigational systems (Llaneras, 2000; Stutts et al., 2001). Tasks that require the direct use of the hands while driving, such as touching a screen or pushing a button on the central console, are major factors in traffic accidents (Klauer, Dingus, Neale, Sudweeks, & Ramsey, 2006). An investigation conducted by the American Automobile Association in 2006, which classified distraction-inducing activities into various types and analysed them according to their weighted contributions to driver distraction, showed that tasks such as turning on or controlling an audio device accounted for the highest proportion of distracting activities (72.5%) (Stutts et al., 2003). Thus, statistical analyses of the causes of traffic accidents indicate a need to consider human factors when designing in-vehicle systems in order to minimize their potential for distraction and the possibility of contributing to an accident.

New types of in-vehicle control systems that support natural interactions between humans and their devices are being developed (Hong & Woo, 2008). In-vehicle system controls can be classified into four types: tactile, touchscreen, speech, and gesture control (Bach, Jaeger, Skov, & Thomassen, 2008). Traditional tactile controls include push buttons and rotary switches, whereas more recent forms of tangible control use touchscreens, which may cause significant visual and biomechanical distractions if they are located away from the driver's line of sight and hand position, for example, on the central console. Research has shown that touchscreen controls cause more severe visual and cognitive distractions than traditional tactile controls (Tsimhoni & Green, 2001; Noy, Lemoine, Klachan, & Burns, 2004; Bellotti, De Gloria, Montanari, Dosio, & Morreale, 2005). Speech control is the most natural form of control and causes little visual and biomechanical distraction because it facilitates hands-free and eye-attention-free interaction; however, it can contribute to auditory distractions (Gellatly, 1997; Barón & Green, 2006). Gesture control is considered an alternative to speech control and has the potential to overcome some of the weaknesses of the latter. As with speech control, gesture control is natural and can reduce visual distractions; however, it may cause biomechanical distractions, and the types of systems using such control are currently limited.

To overcome the problems of gesture control, it is necessary to develop gestures that allow the driver to keep their hands on the steering wheel and their eyes on the road (Werner, 2014). The Geremin system sensed 17 distinct index finger movements away from the wheel as gestures (Endres, Schwartz, & Müller, 2011). The WheelSense system used an ergonomic analysis of hand positions on the steering wheel to evaluate four different wheel grip gestures: two forms of rotation, dragging, and squeezing (Angelini et al., 2013). Several research groups (González, Wobbrock, Chau, Faulring, & Myers, 2007; Döring et al., 2011; Ulrich et al., 2013) have used different variations of thumb-based gestures on a touch-sensitive surface on the steering wheel with small sets of gestures (4 to 19) to allow gesture interactions while holding the steering wheel with both hands. All on-wheel gesture interactions rely on a small number of distinguishable thumb or finger gestures as their main interaction technique. Consequently, they impose several limitations on the user relating to their limitation in number, vague and imprecise meaning, difficulty in remembering, and the lack of passive feedback, i.e. feedback on the current status of the systems and variable values.

Combining gestures with different types of feedback may be a solution to overcome those limitations and could create a powerful and diverse in-vehicle interaction platform. When touchscreens are used as a gesture input, they frequently demand visual feedback, although eyes-free gestures on touchscreens can be facilitated by also providing auditory feedback (Bach et al., 2008), as explored by Angelini et al. (2013). Another common approach for 'eyes on the road' gestures is providing visual feedback with a head-up display (HUD) (Koyama et al., 2014). Numerous studies have illustrated the benefits of HUDs over the head-down displays (HDDs) for the presentation of information related to the operation of the in-vehicle system as well as the vehicle itself (Sojourner & Antin, 1990; Kiefer, 1998; Liu & Wen, 2004; Prinzel & Risser, 2004; Weinberg, Harsham, & Medenica, 2011; Lauber, Follmann, & Butz, 2014; Skrypchuk, Langdon, Sawyer, & Clarkson, 2020). They have identified shorter display–road transition and eye accommodation times (Sojourner & Antin, 1990; Kiefer, 1998) and found the HUD-based interface had a low impact on mental load and scored highest in user satisfaction (Weinberg et al., 2011) in comparison to HDDs. However, there are also problems with HUDs as summarized in Prinzel and Risser (2004), mainly known as cognitive capture, attention capture, or perceptual tunnelling. To upgrade the efficiency and safety, a HUD system must provide drivers with not only large amounts of information from many categories (e.g. route guidance/navigation, traffic signs, cargo/road/vehicle conditions) but also the best way to display this information; important considerations include having a user-friendly system, since a driver's capacity to process this information is a key factor in its acceptance and use (Liu & Wen, 2004).

In this work, we developed a new form of gesture control where a specific number of fingers of the left and/or right hand are opened and closed while the hands remain on the steering wheel – termed on-wheel finger spreading gestures – and combined this with a visual interface displaying the device control menu for audio and air conditioning (A/C) on a HUD. We performed driver-in-the-loop experiments under various driving scenarios to evaluate the effectiveness of our hybrid interface against a traditional tactile interface. A preliminary study was conducted, and its results have been presented (Lee, Yoon, & Shin, 2015) prior to this work. In response to comments from reviewers and attendees at the presentation, we redesigned and conducted the experiment with more participants and expanded its contents significantly. The contributions of the study are as follows:

- We propose a new type of gesture called *on-wheel finger spreading gestures,* which enable the driver to keep their hands on the steering wheel and are easy to perform regardless of the rotation position of the steering wheel.
- We propose a new user interface for in-vehicle devices using on-wheel finger gestures and a HUD for naturalistic input and information display, which can reduce distractions to the driver and enhance response times to avoid dangers in unexpected or hazardous road conditions.
- We verify the effectiveness of the new interface by conducting human-in-the-loop experiments using a driving simulator and comparing the results with those of a traditional interface.
- The results show that the proposed interface was around 20% faster in emergency response than the traditional tactile interface, whereas the performance of the new interface in maintaining vehicle speed and lane did not differ significantly from that of the traditional interface.

This paper is organized as follows. Section 2 surveys related work on gestural or multimodal interfaces for in-vehicle systems. Section 3 introduces our newly proposed interface for audio and A/C systems, which was developed based around on-wheel gesture controls and a HUD. Section 4 describes human-in-the-loop experiments conducted using a driving simulator. Section 5 presents a quantitative comparison and evaluation of the effectiveness between the traditional and proposed interfaces, together with a statistical analysis of the experimental results. Section 6 discusses the main findings from the experiment and the limitations of this study. Section 7 summarizes the results and proposes future work.

## 2. Literature Review

González et al. (2007) stressed the importance of designing gestural interfaces that help the user to maintain 'eyes on the road and hands on the wheel'. They embedded a small touchpad called a StampPad in a steering wheel and evaluated seven methods for selecting from a list of over 3000 street names. Selection speed was measured while being stationary and while driving a simulator. The results showed that the EdgeWrite gestural text entry method was approximately 20 to 50% faster than selection-based text entry or direct list-selection methods. They also showed that a faster driving speed generally resulted in slower selection speed. However, with EdgeWrite, participants were able to maintain their speed and avoid incidents while selecting and driving at the same time. This work was a first attempt to explore and evaluate emerging on-wheel hand gestures adopted in automotive settings. Nonetheless, since this system requires two touchpads to be mounted at the 2 o'clock and 10 o'clock positions, it is difficult to use this interface when the steering wheel is rotated. In contrast, the gestures proposed in our system can be used independently of the rotational position of the wheel.

Bach et al. (2008) presented an approach towards in-vehicle gestural interaction, comparing tactile, touchscreen, and gestural interaction to control a radio. For gestural inputs, a touchscreen mounted on the vertical centre stack was used. They focused on the effects on driving and evaluated three interaction techniques using 16 subjects under two different settings. Their results indicated that, although gestural interaction is slower than touch or haptic interaction, it can reduce distractive eye glances while interacting with the radio. In their work, the gesture interface uses a touchscreen as a drawing canvas, thus the right hand is taken off the steering wheel. On the contrary, our gestures are made while both hands remain on the wheel, which can reduce both biomechanical and visual distractions.

Döring et al. (2011) developed a user-defined set of steering wheel gestures and a working prototype based on a multi-touch steering wheel for gesture identification and compared their application against conventional user interaction with an infotainment system, considering driver distraction. Their results showed that using gestures reduces the visual demand for interaction tasks; however, gestures introduce a similar problem as buttons: scalability. By using gestures that do not need visual attention, the gesture rapidly becomes complicated and harder to remember. The use of touch interaction relating to the displayed content on the screen reduces the benefit of reduced visual attention. This study motivated us to investigate a user interface combining on-wheel gestures with a graphic interface using a HUD. Since Döring et al. used a system with a multi-touch canvas on the steering wheel for touch input and graphical output, the driver's gaze can be taken off the road. To address this problem, we introduce a HUD for graphical output and on-wheel finger gestures for non-contact input.

Pfleging, Schneegass, and Schmidt (2012) proposed a multimodal interaction system combining speech and gesture interfaces, in which voice commands are used to select visible objects or functions while simple touch gestures are used to control these functions. With such an approach, it is simpler to recall voice commands as users see what they need to say. Using a simple touch gesture, the interaction style lowers visual demand and simultaneously provides immediate feedback and an easy means for undoing unwanted actions. A set of user-elicited gestures and common voice commands were determined in a user-centred system design process. In an experiment with 16 participants, Pfleging et al. explored the impact of this form of multimodal interaction on driving performance against a baseline using physical buttons. Their results indicate that the use of speech and gestures is slower than using buttons but results in similar driving performance. Users commented in a questionnaire that the visual demand was lower when using speech and gestures. The overall distraction of this multimodal interaction is comparable to the current interaction approach but offers greater flexibility. In particular, voice controls can cut down on the number of clicks required in multi-levels of a menu structure; however, naming objects and functions could be difficult, and the users may be required to learn and remember them. To avoid these issues, we do not adopt voice controls but instead use HUD and audio feedbacks to confirm that a selection was performed correctly.

Angelini et al. (2013) presented a novel opportunistic paradigm for in-vehicle gesture recognition allowing the use of two or more subsystems in a synergistic manner. In order to segment and recognize micro-gestures performed by the user on the steering wheel, they combined a wearable approach based on the electromyography of the user's forearm muscles with an environmental approach based on pressure sensors directly integrated into the steering wheel. Several fusion methods and gesture segmentation strategies were presented and analysed, whereupon a prototype was developed and evaluated with data from nine subjects. Their results showed that the proposed opportunistic system performs equal to or better than each standalone subsystem while increasing interaction possibilities. The micro-gesture interface using pressure sensors helps the user to keep their attention on the road and their hands on the steering wheel. However, attaching electromyography (EMG) sensors to the user's arms is not practical. Our current method uses the same advantageous aspects of on-wheel micro-gestures. Although we currently use data gloves for recognizing finger gestures, we plan to replace them with the fusion of pressure and vision sensors in the future.

Angelini et al. (2014) presented the results of a user elicitation study for gestures performed on the surface of the steering wheel. Forty participants elicited a total of 240 gestures. The study provided useful information about gestures that users are likely to expect in an in-vehicle gestural interface. This information could help in the design of steering wheels that detect gestures on their surfaces. Thus, technologies based on proximity, capacitive, or pressure sensors can be used to provide an interaction compliant with the 'eyes on the road, hands on the wheel' paradigm, if coupled with proper feedback such as a HUD. We adopted the results of this study to combine our gestural interface with a HUD for visual feedback.

Koyama et al. (2014) developed a multi-touch steering wheel where touch positions correspond to different operating positions in order to control the information offered by in-car tertiary applications. Thereby, drivers could operate applications at any position on the steering wheel. One hundred and twenty infrared sensors were embedded in the steering wheel and the system was trained using a Support Vector Machine algorithm

Journal of Computational Design and Engineering, 2020, 7(6), 700–721 | 703

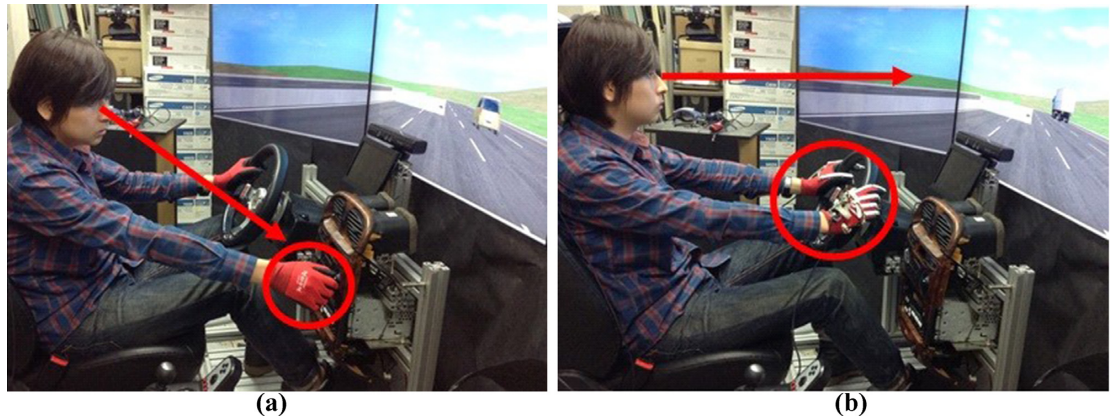

**Figure 1:** Head and eye movements and hand positions during in-vehicle device control (Lee et al., 2015): (a) a traditional interface using a tactile interface on the central console and (b) a new interface using on-wheel gestures and a HUD.

| Function selection gesture | Detailed function selection (numerical control) gestures | | | | Function selection gesture | Device turn-off gesture |
|---|---|---|---|---|---|---|
| | Channel increase | Channel decrease | Volume increase | Volume decrease | | |
| Left-hand index raise | Left-hand index raise | Left-hand thumb twist | Right-hand index raise | Right-hand thumb twist | Left-hand index raise | Left-hand spread |

**Figure 2:** A series of on-wheel gestures to control radio and audio devices proposed by Kang (2012).

to recognize different hand gestures (flick, click, tap, stroke, and twist). Additionally, navigation and audio applications were implemented on the proposed interface. Koyama et al. conducted a user study for navigation application and found an average flick recognition rate of about 92%. However, to select a function, the driver had to browse the menu using a flicking gesture until reaching the desired function. This selection process may cause visual and cognitive distractions. The menu structure of an application thus needs to be designed taking into consideration the cognitive workload and visual demand.

Riener (2012) surveyed in-vehicle gestural interfaces and summarized the advantages and disadvantages of various finger- and hand gesture recognition systems. Ha and Ko (2015) proposed a vision-based shadow-gesture recognition method for interactive projection systems, which separated the shadow area, isolated hand shadows to distinguish hand gestures, and tracked the fingertips using an optical flow algorithm. Rempel, Camilleri, and Lee (2004) studied discomfort and fatigue resulting from hand gestures associated with characters and words used by professional sign language interpreters. Wachs, Kölsch, Stern, and Edan (2011) extensively evaluated the design of hand gestures and recommended four guidelines for future hand gesture interfaces to increase the likelihood of their widespread commercial or social acceptance: validation, user independence, usability criteria, and qualitative/quantitative assessment.

Weinberg et al. (2011) evaluated the usability of a HUD for selection from choice lists in car. The experiments on a driving simulator showed that the HUD had a low impact on mental load and scored highest in user satisfaction among three output system variants for in-vehicle systems: a HDD, a HUD, and an auditory display. Lauber et al. (2014) presented the 'What You

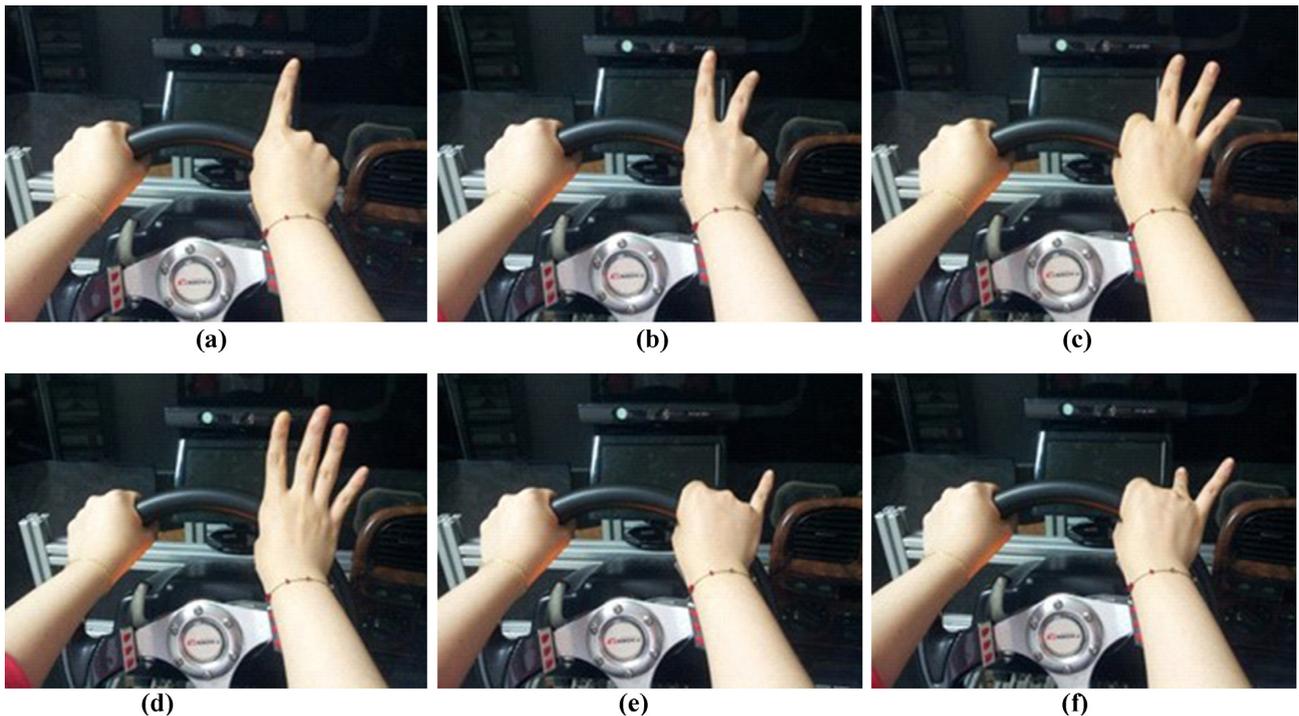

**Figure 3:** Spreading the fingers to indicate different numbers using on-wheel finger spreading gestures: indicating (a) one by spreading the index finger; (b) two by spreading the index and middle fingers; (c) three by spreading the index, middle, and ring fingers; (d) four by spreading four fingers excluding the thumb; (e) one by spreading the little finger; and (f) two by spreading the ring and little fingers.

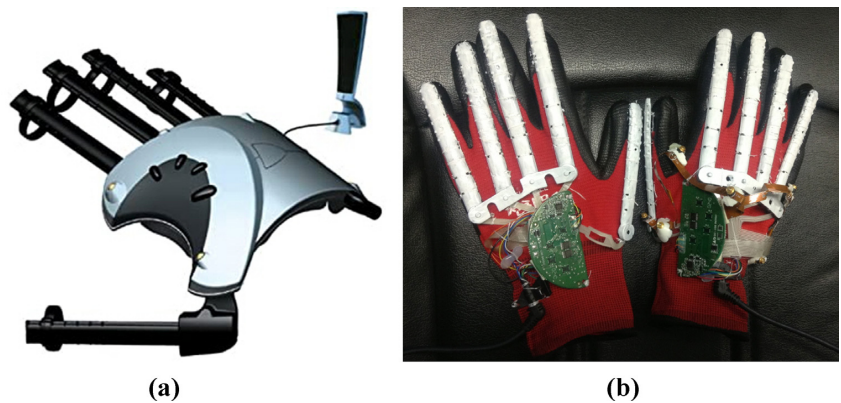

**Figure 4:** Essential Reality P5 Gaming Glove: (a) original right-hand glove and (b) customized right- and left-hand gloves.

See Is What You Touch' (WYSIWYT) technique for touchscreen interaction that no longer requires any direct visual attention on the touchscreen itself. Instead, its content as well as a representation of the user's finger is displayed in the HUD. This creates a shorter distance between the display location and the road scene, which in turn allows beneficial gaze behaviour. Instead of having to fully avert the eyes from the road, the driver can switch their focus back and forth during interaction. They have combined this approach with pointing gestures and introduce several variations of the WYSIWYT technique, some of which allow users to even interact with the touchscreen without actually touching it.

Shim and Lee (2016) proposed an in-vehicle spatial gesture control combined with a HUD to reduce visual distractions for drivers. They selected the controls used most frequently for audio and A/C devices in the central console and implemented these controls using HUD menus and a hand-motion recognition system. Recently, we proposed a new driver interface that replaces spatial gesture control with on-wheel gesture control in which the driver spreads and closes the fingers while the hands remain on the steering wheel (Lee et al., 2015). We compared our proposed system with a traditional interface using tactile control and a HDD using quantitative and objective measurements as well as qualitative and subjective evaluations. However, our study had limitations not only in the scenarios and participants in the experiment but also in objective measurements. Therefore, in response to comments from reviewers and attendees at its presentation, we redesigned the system and conducted the experiment with more participants and expanded its contents significantly. We also considered the large volume of research on the application of virtual or augmented reality to a user interface, which may be applied to user interfaces in the future (Takahashi et al., 2018; Fukuda Yokoi, Yabuki, & Motamedi, 2019; Son, Jang, & Choi, 2019; Sun, Hu, & Xu, 2019).

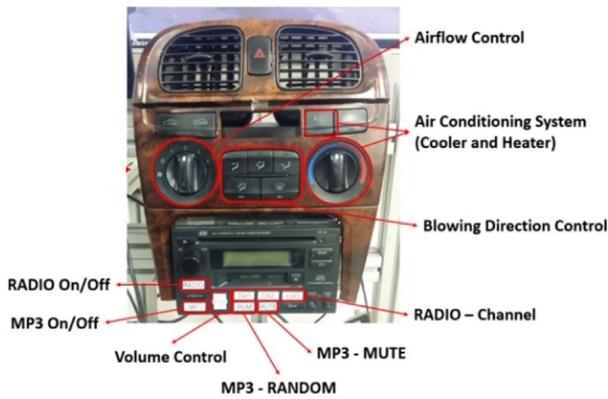

**Figure 5:** Audio and A/C controls of a Hyundai Sonata on the central console of the driving simulator.

## 3. Design of User Interface Combined with On-wheel Finger Gestures and a HUD

### 3.1. On-wheel finger spreading gestures

Conventional in-vehicle systems are operated via tactile interactions with buttons on the lower part of the unit or by turning the volume controller knob. Thus, the user generally experiences tactile feedback when pressing buttons or turning knobs. Visual distraction occurs when the driver's vision shifts to the central console, which is located to the lower right of the steering wheel, in order to execute tactile control of the audio and A/C systems, as shown in Fig. 1a. In addition, biomechanical distraction occurs when the driver's hand moves away from the steering wheel. To address these problems, we developed a new form of gesture control in which the fingers are allowed to move while the hands remain on the steering wheel, as shown in Fig. 1b, together with a new interface that displays the device control menu on a HUD to minimize visual distractions.

The idea of our finger gesture control was inspired by Kang (2012). In his thesis, through user survey, he extracted intuitive and useful on-wheel gestures and one-hand gestures near the area of the central console for controlling in-vehicle devices such as the radio, audio, A/C, and heater. Then, these candidate gestures were evaluated by design experts to select examples for inclusion in the final set-up. Next, through the one-to-one interviews of an expert group, the gestures were mapped to specific operations of in-vehicle devices. Following an interview with a group of experts, the index raise and thumb twist gestures were selected as the simplest and most intuitive for controlling in-vehicle devices. The left-hand index raise and left-hand thumb twist gestures were also chosen to increase and decrease channels in the radio and audio, and the temperatures in the heater and air conditioner, as shown in the gesture scenario presented in Fig. 2. The right-hand index raise and right-hand thumb twist gestures were chosen to increase or decrease the volume of the radio and audio, and the airflow of the heater and air conditioner. The hand spreading gesture was selected to turn off devices since this gesture is clearly distinct from the others. The thumb movement, however, is not well recognized by vision-based gesture recognition systems. To overcome this, it is necessary to mount small touchpads on the steering wheel, typically at the 2 o'clock and 10 o'clock positions, as proposed by González et al. (2007). However, it is difficult to use touchpads when the steering wheel is rotated. Moreover, these proposed gestures have not been evaluated and verified through human-in-the-loop experiments using a driving simulator.

To improve Kang's on-wheel gestures, we adopted only the index raise gestures and combine these gestures with a HUD. In this paper, we refer to spreading and closing of the fingers while the hands remain on the steering wheel as *on-wheel finger spreading gestures*, or simply as *on-wheel finger gestures*. As shown in Fig. 3, such on-wheel finger gestures can be applied to operate a specific switch according to the number of spreading fingers while the hands remain on the steering wheel. The reason for choosing the number of fingers as the output of the gestures is that there are up to 25 (= 5 × 5) choices from the combination of the spread finger of the right and left hands. When a multimodal interface is designed by combining gestures and a display, many such choices give freedom to design the menus of the graphical user interface (GUI). In this work, gestures were used to select menus on the HUD, but we did not distinguish which fingers were spread because some finger gestures may be physically stressful or preferred depending on individual users. On-wheel finger gestures facilitate immediate responses to danger because both hands remain on the steering wheel. Moreover, the finger that is spread does not need to be a specific finger, thereby avoiding differences in finger spreading that may occur from cultural differences, personal preference, or physical stress (Rempel et al., 2004). A human–machine integrated design and analysis framework based on motion synthesis and biomechanical analysis can be applied for quantitative discomfort evaluation of finger and hand gestures (Choi & Lee, 2015; Lee, Jung, Lee, & Lee, 2017).

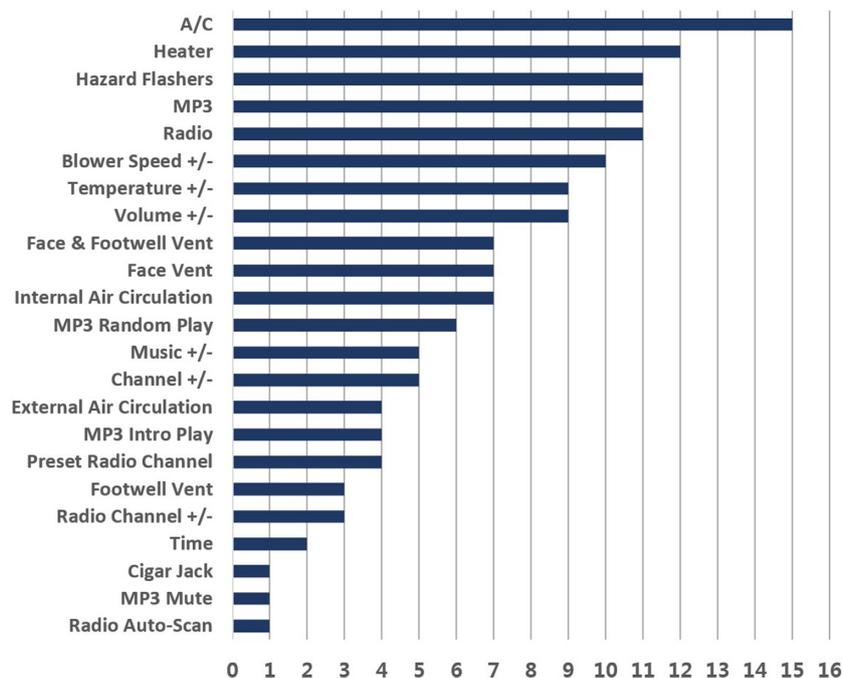

**Figure 6:** Results of the questionnaire survey used to determine the most frequently used switches on the central console.

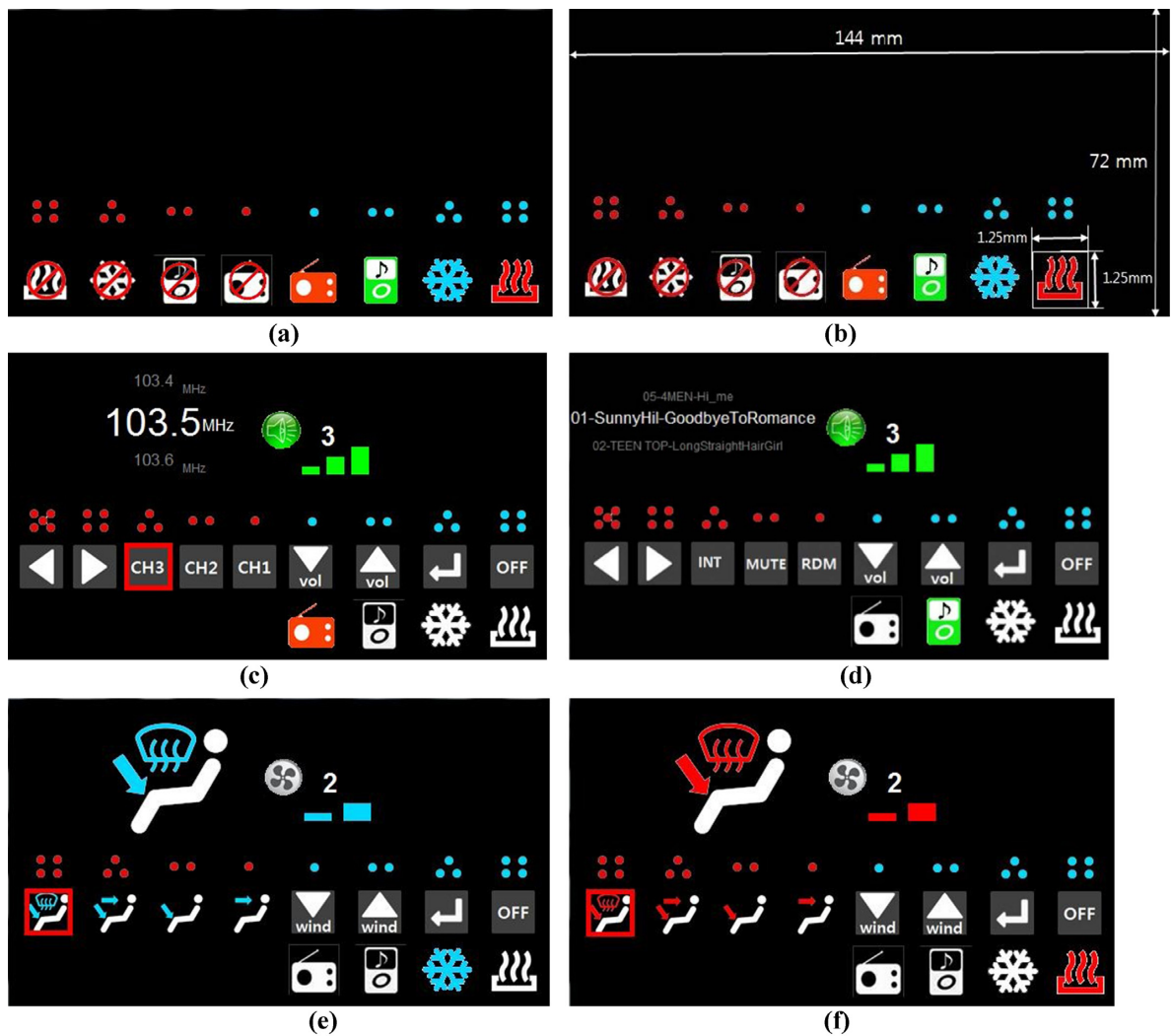

**Figure 7:** HUD graphical user interface: (a) initial menu; (b) radio; (c) MP3; (d) A/C; and (e) heater control.

In general, gestures were captured in images by a camera or signals by motion sensors, which were then recognized using pattern-recognition processes. The aim of this study was not gestured recognition itself; thus, a data glove, and not computer vision technology, was used to capture finger movements. On-wheel gestures were captured using an Essential Reality P5 Gaming Glove. As shown in Fig. 4, the glove was disassembled and its components were attached to a cotton glove, which allowed the glove to be worn easily and increased the rate of recognition of finger bending. Furthermore, since only a right-hand glove was available, the components from a disassembled right-hand glove were used to produce a left-hand glove. This glove obtains finger-bend information, and coordinates information along the X, Y, and Z axes, when used in conjunction with an infrared tracking unit. The finger sensors provide five independent finger measurements at a 60 Hz refresh rate and 0.5° resolution. The range of finger bending is 0 (completely closed) to 63 (completely opened). In the system, the threshold was set to 30, meaning that the system recognizes a finger as straightened if its bend value is greater than the threshold. We tested our system using a set of 10 selected natural gestures, performed multiple times by 20 different persons. Our system is able to recognize finger spreads correctly in more than 99% of instances with a false positive rate lower than 0.5%.

### 3.2. Graphic user interface on a HUD

The central console of the Hyundai EF Sonata used in this study comprised an A/C control at the top and an audio control at the bottom, as shown in Fig. 5. These controls use push buttons and rotary switches. The central console was located at the centre of the cockpit, to the lower right of the steering wheel. The central console contains various switches for the audio and A/C systems. Twenty-four participants were asked to identify the switches that they use most frequently, indicating that the ON/OFF switches for the A/C (cooler and heater) and audio systems (radio and MP3 player), fan, volume, and thermostat were used most frequently, as shown in Fig. 6. Therefore, in the present study, we applied on-wheel gesture controls to these switches as hotkeys, and all switches on the central console can be used in parallel. However, all switches on the central console can be accommodated in the graphic interface menu, although this results in increasing the depth of the menu, which may cause cognitive distractions.

When creating user interfaces, most designers follow interface design principles that represent high-level concepts used to guide software design. Several fundamental principles, such as '10 Usability Heuristics for UI Design' (Nielsen, 1995), 'The Eight Golden Rules of Interface Design' (Shneiderman, 1998), and 'First Principles of Interaction Design' (Tognazzini, 2014), are commonly used. Among these UI design principles, we selected and applied the following: (i) Reduce Cognitive Load – our multimodal interface promoted recognition in UI by making information and functionality visible. In addition, we added pips above icons to reduce the user's cognitive use to recall the finger count. (ii) Make UI Consistent – our UI requires consistent sequences of actions in similar situations. The heater and cooler share the same UI, as do the radio and MP3 player. (iii) Offer Informative Feedback – the UI offers visual feedback together with verbal and sound feedback.

In this study, both hands were used, and the left and right hands were distinguished during on-wheel gesturing. Thus, a graphical interface was developed and displayed on the windshield, as shown in Figs 7 and 8, based on experience in GUI designs of gauge clusters (Eom & Lee, 2015; Lee & Ahn, 2015). The initial menu shows the arrangement of the ON/OFF switches for the radio, MP3, A/C, and heater, as shown in Fig. 7a. The switches that turn on these devices are shown on the right, and the switches that turn them off are shown on the left. The pips above each switch represent the number of fingers that need to be spread. Red indicates the left hand, and blue indicates the right hand. After selecting a device from the initial menu, the user is directed to the submenu for controlling the selected device. Figure 7c shows the radio control menu. The selection menus for radio channels 1, 2, and 3, and for selecting one channel up and one channel down are arranged in order from the centre to the left, whereas the menus for selecting one level of volume up or down, returning to the previous menu, and turning off the menu are arranged in order from the centre to the right. The MP3 control menu is shown in Fig. 7d. The controls are arranged from the centre to the left, i.e. random play, mute, and intro play (to browse and preview each file for 10 s), which can be toggle-operated, and the right-side menu is the same as that for the radio. The menus for the A/C and heater are shown in Fig. 7e and f, respectively. The four menus on the left indicate the air outlet directions, whereas the menus on the right indicate one level up and one level down for the fan, one level up for the menu, and control off. We also provide hotkeys to turn the system on and off and to cancel the current input. To turn the system on and off, the user must spread two hands. To cancel the current input and return to the previous state, the user must spread all the fingers of the right hand. To go to the upmost menu, the user spreads the index fingers of both hands.

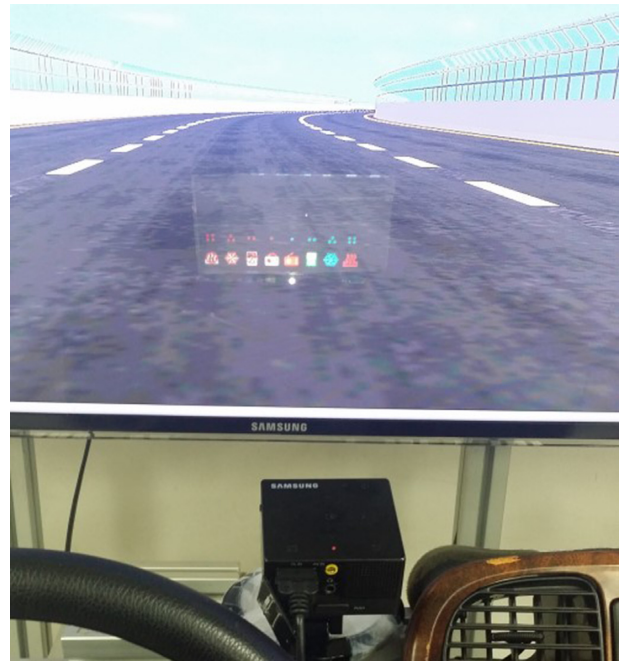

**Figure 8:** A HUD menu displayed on the monitor.

Since gesture recognition systems have inherent accuracy limitation, it is very important to give feedback to the user in order to make the recognition process less opaque. For this reason, we upgraded the system to provide audio feedback on the completion of a gesture. When a device is selected on the top menu, the system gives verbal feedback, i.e. the name of the device is heard. When an operational function is selected on a submenu, a simple electronic sound is generated, or the sound of an audio device is heard. We chose to implement only the simplest audio feedback, retaining more sophisticated audio feedback for future work.

### 3.3. User interface with a HUD and on-wheel finger spreading gestures

The overall architecture of the system is illustrated in Fig. 9. Finger opening and closing motions are captured by the bend sensors of data gloves. The finger-bend values are transferred to the gesture recognition module, in which a finger is determined to 'spread' if passing the threshold value. According to the number of opened fingers of the right and left hands, a specific icon representing the device or control is selected from the current menu on the HUD. Finally, the selected control of the device is executed.

An example of controlling radio volume using this new interface is shown in Fig. 10. The radio is selected when one finger is spread, as shown in the first image in the figure. Next, spreading two fingers on the right-hand increases the volume. Four fingers are then spread on the right hand to select the OFF state, leave the submenu and return to the initial menu screen.

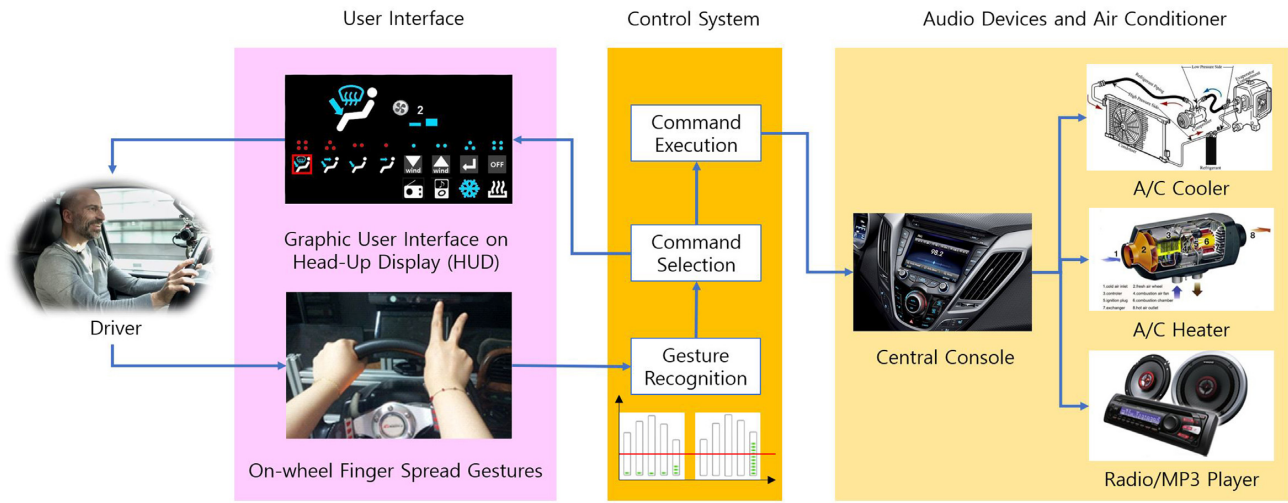

Figure 9: Overall architecture of the system equipped with the proposed user interface combined with on-wheel finger spreading gestures and a HUD.

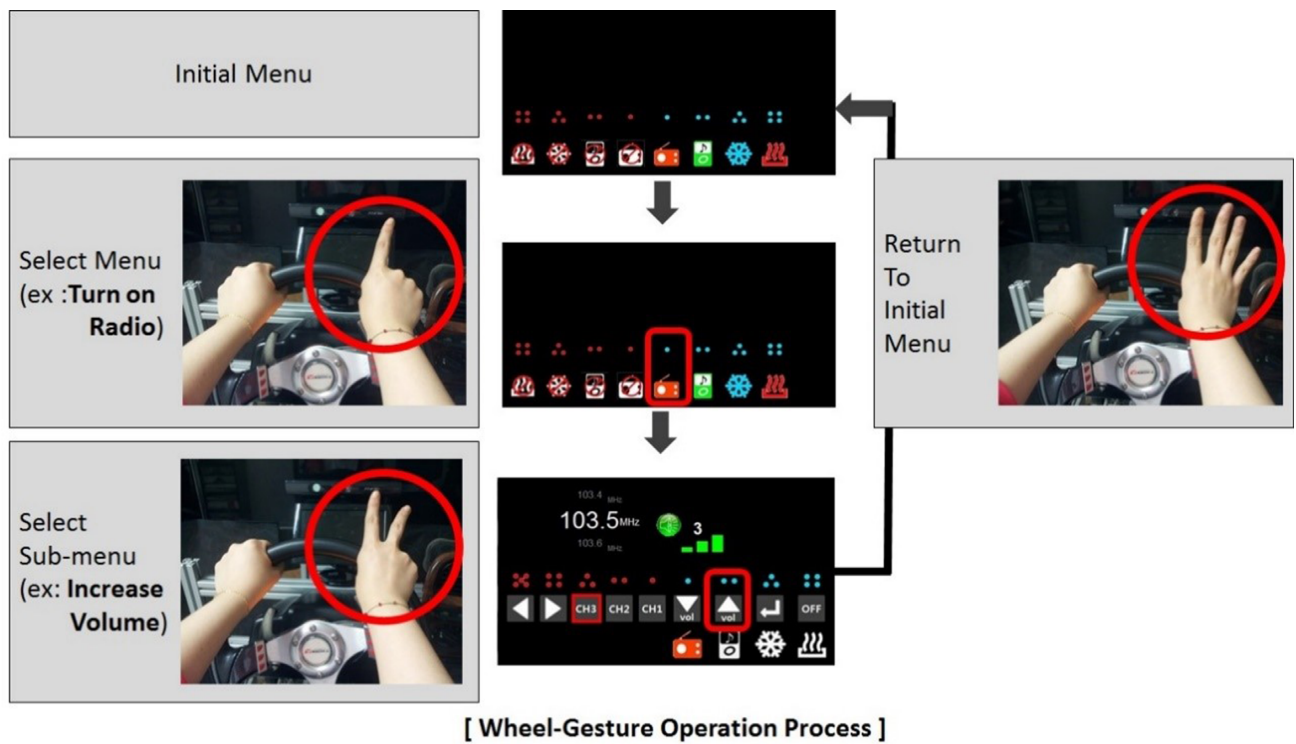

Figure 10: Example of radio control using the new interface based around on-wheel finger spreading gestures.

## 4. Method

We compared a traditional interface equipped with tactile control and a HDD with a new interface equipped with non-tactile on-wheel gesture control and a HUD through human-in-the-loop experiments. We investigated their performance in terms of primary and secondary tasks, as well as visual and biomechanical distractions. Here, the primary task indicates how well the driving task is executed, and its performance is evaluated by observing how well the driver maintained the vehicle at a specific speed and in a specific lane. Secondary tasks involve the control of audio and A/C systems via a central console. Performance was evaluated by measuring how accurately and rapidly the driver executed a given secondary task while driving. Finally, visual distractions were evaluated by determining the extent to which the sightline was not focused on the road, whereas biomechanical distractions were evaluated by determining how much the body deviated from the driving position. The movements of the head and the user's line of sight were evaluated using head and eye tracking systems.

### 4.1. Subjects

Using poster advertisements and internet bulletin boards, we recruited subjects who satisfied the following requirements: possessing a driver's licence, at least 12 months of actual driving experience, and good health with no illnesses, regardless of gender and age. In total, 32 subjects participated in the experiment, i.e. 28 males and 4 females with an average age of 24.3 (SD = 2.1) and an average driving experience of 40.5 (SD = 15.7) months.

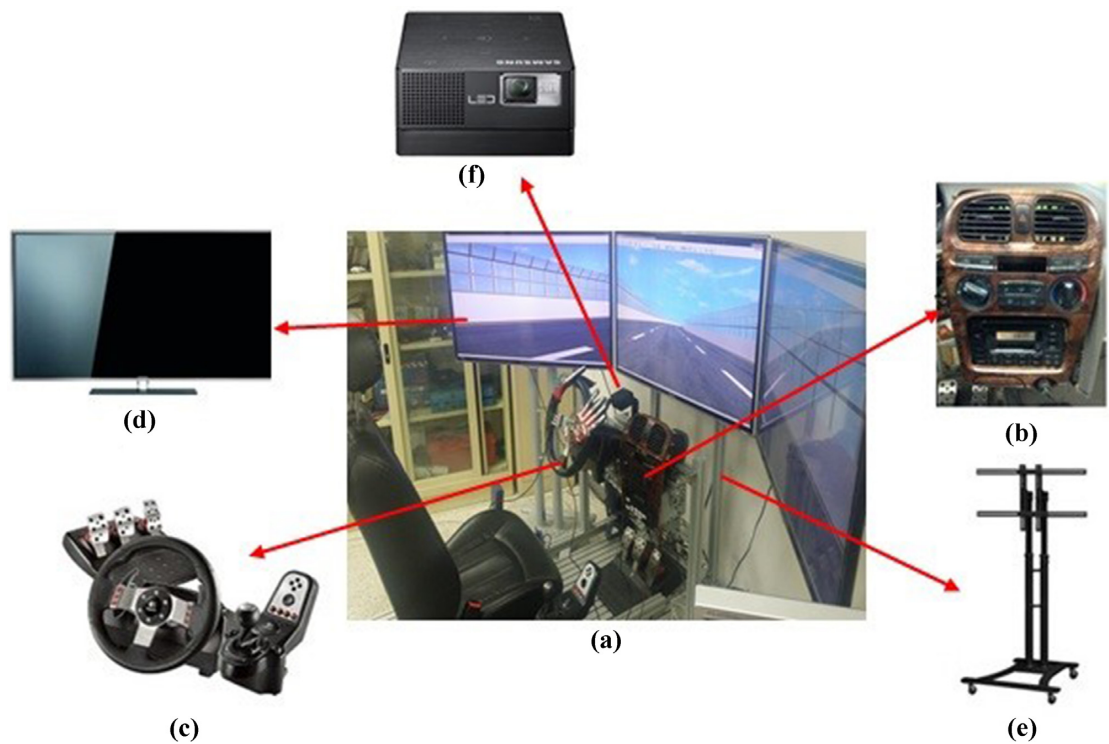

**Figure 11:** Experimental equipment: (a) vehicle mock-up; (b) central console from a 1997 Hyundai EF Sonata; (c) Logitech G27 Racing Wheel; (d) Samsung 40-inch LED TV; (e) Media First D-8600A monitor holder; and (f) Samsung SP-H03 Pico mini-beam projector.

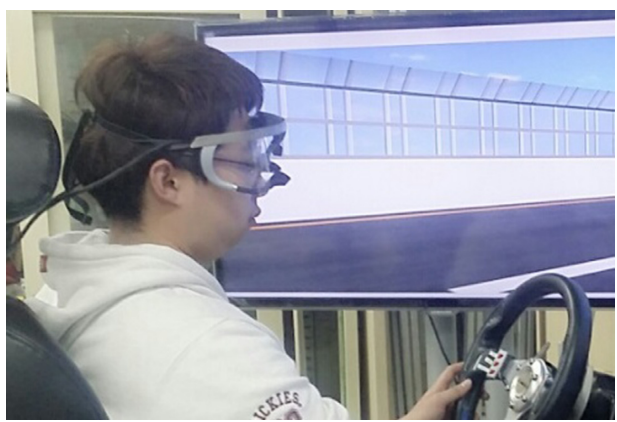

**Figure 12:** Tracking eye movements using Ergoneers Dikablis Professional glasses.

### 4.2. Apparatus

#### 4.2.1. Driving simulator
To provide a simulated driving environment, we used UC-win/Road (2019), a 3D road and traffic modelling and driving simulation software system developed by FORUM8. As shown in Fig. 11, the vehicle mock-up comprised a frame of 980 mm × 1580 mm × 2000 mm (width × length × depth), which was constructed with aluminium extrusion profiles. The mock-up was installed with a central console and a seat from a Hyundai Sonata. The steering wheel, brake and accelerator pedals were from a Logitech G27 Racing Wheel. The dual-motor force feedback of the 11-inch steering wheel equipped with a spiral gear allowed the driver's hands to feel changes in vehicle weight and sliding of the tires. The driver could make 2.5 turns lock to lock. The display equipment comprised three Samsung 40-inch LED TV monitors to provide a field of view of 170°. A mobile D-8619A monitor holder from Media First Co., able to support a monitor of up to 55 inches and provide vertical angle control within 25° and height control within a range of 750–1150 mm, was used to hold the monitors. As a HUD, a Samsung SP-H03 Pico mini-beam projector was installed between the steering wheel and the 40-inch monitor at the front. The projector is palm-sized and delivers 30 ANSI lumens for WVGA (854 × 480) resolution.

#### 4.2.2. Eye tracking glasses and head tracking device
To track the driver's eye movements, we used Ergoneers Dikablis Professional glasses (Fig. 12), which are binocular at a tracking frequency of 60 Hz (per eye) and fit over all types of glasses. The scene is recorded at a resolution of 1920 × 1080 pixels (full HD) and a frame rate of 30 fps using a camera with an aperture angle between 40° and 90° (Ergoneer Dikablis Glasses, 2016). The glasses are automatically corrected for phase differences between the camera and eye and can be used to monitor the driver's gaze in real time. Experiments can also be easily conducted because a separate marker is not needed to calculate the device's position. The glasses used can record video information related to the driver's field of view and gaze. A post-processor can be used to specify a certain region in the captured image and obtain the statistics regarding visual fixation time and frequency for this region.

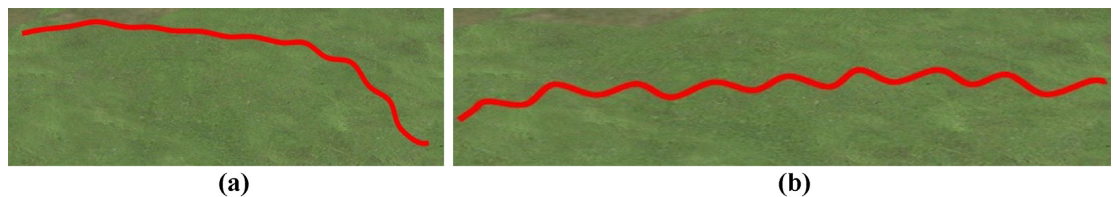

**Figure 13:** Road environment modelling: (a) first road with easy conditions and (b) second road with more curved sections.

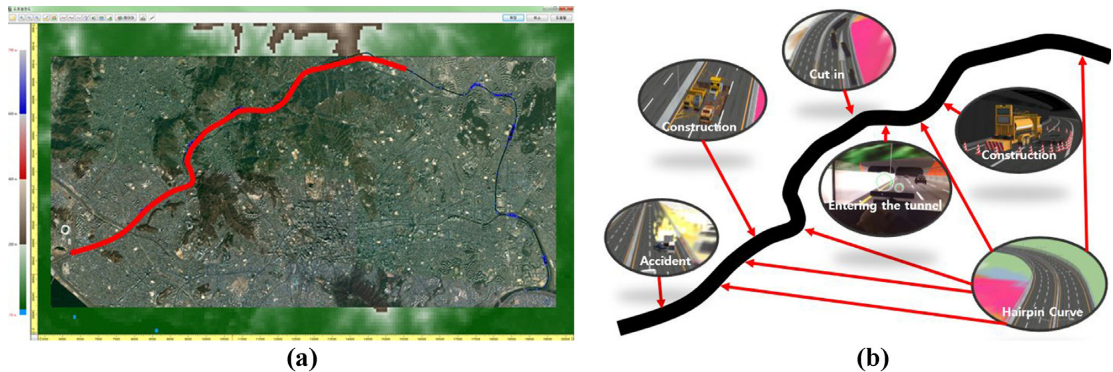

**Figure 14:** Road environment configuration: (a) the third road, based on a section of the urban highway in Seoul (Lee et al., 2015) and (b) hazardous conditions on this road.

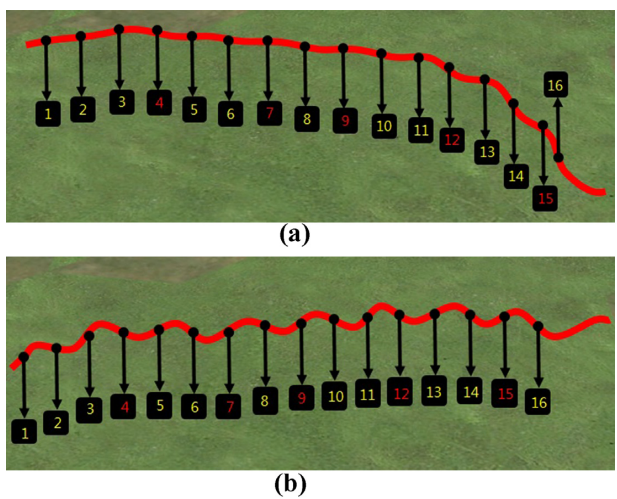

**Figure 15:** Road task locations: (a) the first easy road and (b) the second difficult road.

Microsoft Kinect was used to construct a head tracking system capable of capturing head rotations. We used the Single Face programme from Kinect's Face Tracking Visualization SDK, which tracks the driver's face to show an avatar in the left window and the polygon-masked face of the driver in the right window, both in real time. The avatar drawing code was modified to allow the yaw, pitch, and roll angles of the driver's face to be recorded at frequencies of 60 Hz.

*4.2.3. Road environment*
UC-win/Road, a 3D urban visualization and transport modelling software system, was used in the experiment to construct the road environment, which was composed of two roads as shown in Fig. 13. The first road is easy to drive with gently curved sections. The second road is challenging to drive due to having more curved sections. Each road is 4.9 km in length with six lanes (three in each direction). Both roads included multiple curved sections and unexpected conditions, such as accidents. There were no cars on the road except for those involved in accident conditions.

We also modelled a section of the urban highway in Seoul, as shown in Fig. 14. This third road is 11.55 km in length with six lanes (three in each direction) and has two tunnels and multiple curved sections. It also includes unexpected conditions, such as accidents or roadwork sites.

### 4.3. Secondary task scenario

For the first and second roads, the participants were asked to control the audio and A/C systems at 16 locations while driving, as shown in Fig. 15. During the experiment, the subjects were asked to maintain the specified reference speed of 80 km/h and were asked to drive in the second lane throughout the experiment. Table 1 summarizes the locations at which the task commands were given, together with the road conditions and vehicle lanes at these locations and the details for each of the 16 tasks. Road conditions were classified into two main types: normal conditions with no danger on the road and hazardous conditions with accidents. Among the 16 tasks shown in Table 1, 11 tasks were conducted under normal road conditions, whereas the remaining 5 tasks were conducted under hazardous conditions. The subjects performed the 16 tasks on each road for each user interface. Here, the 'Levels' in the last column refer to how many levels the driver should increase or decrease the volume of an audio device or the airflow strength of an air conditioner.

For the third road, the subjects controlled the audio and A/C systems at 15 locations while driving 11.55 km from the Seongsan Ramp to the Kookmin University Ramp, as shown in Fig. 16. In these experiments, subjects were asked to maintain a specified reference speed of 70 km/h. They were asked to drive from the starting point of the road using the second lane in the 300–3480 m section, the first lane in the 3850–7350 m section, and the second lane in the 7300–11 550 m section. Table 2 summarizes the locations at which task commands were given, the road conditions and vehicle lanes at these locations, and the details for each of the 15 tasks. The road conditions were classified into two main types: normal conditions with no danger on a straight road; and hazardous conditions with sharp curves, accidents, road construction, cuttings, or tunnels. Among the 15 tasks shown in Table 2, 8 were performed under normal road sections and the remaining 7 tasks were conducted along hazardous sections of the road. t

**Table 1:** Tasks executed by the subjects on the first and second road.

| Task no. | Location | Road condition | Task details | | | |
|---|---|---|---|---|---|---|
| | | | Device | Feature | Control | Levels |
| 1 | 100 m | Normal | Radio | Turn | On | |
| 2 | 400 m | Normal | Radio | Volume | Up | 2 |
| 3 | 700 m | Normal | A/C | Turn | On | |
| 4 | 1000 m | Accident | A/C | Top | Select | |
| 5 | 1300 m | Normal | Heater | Turn | On | |
| 6 | 1600 m | Normal | Heater | Airflow | Down | 1 |
| 7 | 1900 m | Accident | MP3 | Turn | On | |
| 8 | 2200 m | Normal | MP3 | MUTE | Select | |
| 9 | 2500 m | Accident | Heater | Turn | On | |
| 10 | 2800 m | Normal | Heater | Bottom | Select | |
| 11 | 3100 m | Normal | MP3 | Turn | On | |
| 12 | 3400 m | Accident | MP3 | Volume | Up | 1 |
| 13 | 3700 m | Normal | Radio | Turn | On | |
| 14 | 4000 m | Normal | Radio | CH1 | Select | |
| 15 | 4300 m | Accident | A/C | Turn | On | |
| 16 | 4600 m | Normal | A/C | Airflow | Up | 2 |

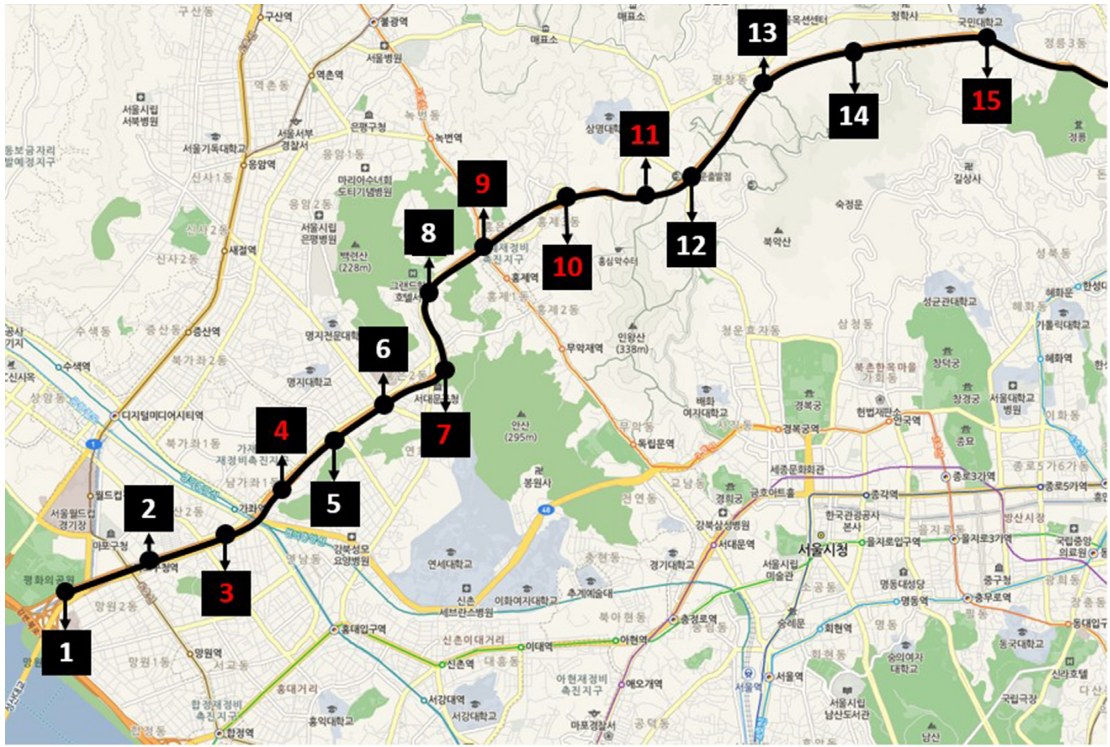

**Figure 16:** Task locations on the third road, representing a section of the urban highway in Seoul.

### 4.4. Procedure

Before the experiment, the subjects were informed about the aims of the study and the methods employed and gave their written consent to participate. They practised driving in the simulator to familiarize themselves with the environment and exercised on-wheel gesture control and traditional push-button and rotary-switch controls for 50 min. To explore their learning rates, they undertook a test after practising 32 tasks for each interface. The average task completion times for each test were recorded for subsequent analysis.

Table 2: Tasks executed by the subjects on the third road.

| Task no. | Location | Road condition | Task details | | | |
|---|---|---|---|---|---|---|
| | | | Device | Feature | Control | Levels |
| 1 | 900 m | Normal | Radio | Volume | Up | 2 |
| 2 | 1270 m | Normal | MP3 | Volume | Down | 3 |
| 3 | 1900 m | Sharp curve | A/C | Airflow | Up | 2 |
| 4 | 2900 m | Sharp curve | Heater | Airflow | Down | 1 |
| 5 | 3480 m | Normal | Radio | CH1 | Select | |
| 6 | 3850 m | Normal | MP3 | Mute | Select | |
| 7 | 4230 m | Sharp curve | A/C | Top | Select | |
| 8 | 4390 m | Normal | Radio | Volume | Down | 3 |
| 9 | 5310 m | Cut In | MP3 | Volume | Up | 2 |
| 10 | 5970 m | Tunnel | Heater | Airflow | Up | 3 |
| 11 | 6700 m | Sharp curve | A/C | Airflow | Down | 2 |
| 12 | 7350 m | Normal | Radio | CH2 | Select | |
| 13 | 9120 m | Normal | MP3 | Random | Select | |
| 14 | 10 530 m | Normal | MP3 | Volume | Up | 5 |
| 15 | 11 110 m | Sharp curve | Heater | Bottom | Select | |

Table 3: Dependent variables used in the experiments.

| Class | Sub-class | Variable | Unit |
|---|---|---|---|
| Objective measurement | Primary task performance | Speed deviation | km/h |
| | | Lateral distance deviation | % |
| | | Brake response time | sec |
| | Secondary task performance | Task completion time | sec |
| | Eye movement | AOI attention ratio | % |
| | | Mean fixation duration | sec |
| | | Horizontal eye activity | pixel |
| | | Vertical eye activity | pixel |
| | Head movement | Head's yaw, pitch, roll | degree |
| Subjective evaluation | Primary task performance | Maintaining speed | 1–5 |
| | | Maintaining lane | 1–5 |
| | Secondary task performance | Performing secondary task | 1–5 |
| | | Confirming setting | 1–5 |
| | | Looking at road | 1–5 |

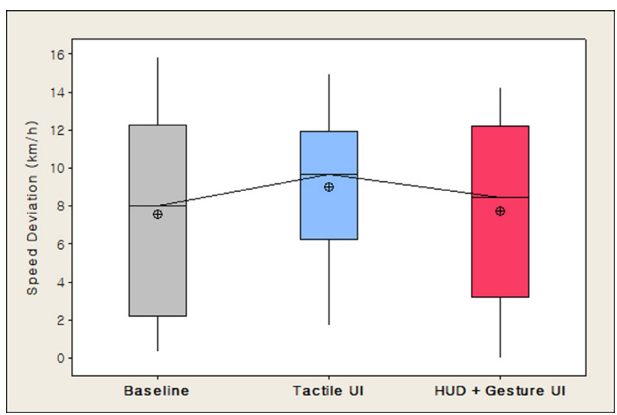

Figure 17: Speed deviation effects for the three roads on which the subjects were tested.

To minimize the effects of training under specific experimental conditions, the subjects were divided into two groups, each of which accessed the two interfaces used in the experiment in a different order. The 32 subjects were divided into two groups of 16 subjects; the first group used the tactile interface first, whereas the second group used the newly proposed multimodal interface first. The tasks for the given scenario were completed within approximately 4 min using each interface. At the location for a specific task, a command was broadcast from a speaker, at which point the subjects began executing the task. Immediately after completing the task, the subjects were required to verbally report 'end' to the experimenter. After the experiment, the subjects were required to complete a questionnaire regarding the experiment. The experiment lasted approximately 3.5 h for each subject.

### 4.5. Variables

The independent variables used in this study were categorical. Two different levels represented the tactile and newly proposed visual and gestural interfaces for the devices on the central console. As shown in Table 3, the dependent variables were classified into two main types: objective measurements and subjective evaluations. These variables were designed according to the approach presented in (Bach, Jæger, Skov, & Thomassen, 2009). Objective measurement variables are values related to the performance of the primary driving task, i.e. speed and lane-keeping capability. The values related to the performance of the secondary task include task completion rate and task completion time for in-vehicle system control, while values related to distraction were obtained from measurements of the driver's eye movements. Subjective evaluation variables comprise values obtained from the questionnaire responses regarding the level of difficulty in maintaining speed and lane, task completion, setting confirmation, and looking forward (i.e. to the road) while operating the device.

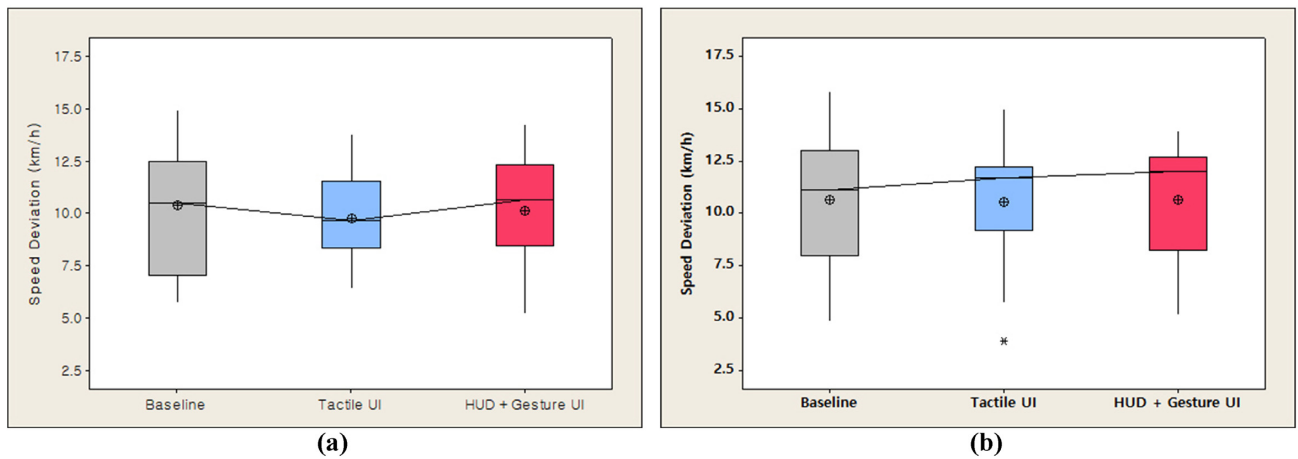

**Figure 18:** Speed deviation effects: (a) on the first easy road and (b) on the second challenging road.

## 5. Results

We established and tested the following null hypothesis: that there is no difference in performance between the tactile and the newly proposed multimodal interfaces, together with an alternative hypothesis, i.e. that there is a difference between them, where one is superior to the other. A repeated measures ANOVA (ANalysis Of VAriance) was performed to analyse the effect of each variable of objective measurement. A Wilcoxon signed-rank test was used for subjective evaluation through a questionnaire survey. The significance level (or $\alpha$ level) was set at 5%, thus the null hypothesis was rejected, and the alternative hypothesis accepted, if $P < 0.05$. Our results are depicted using box plots with mean symbols and median connect lines.

### 5.1. Speed deviation effects

The subjects were required to drive at a constant speed during the experiment. Speed deviation was therefore used to evaluate the effects of conducting secondary tasks on driver performance. Speed deviation was computed as the root-mean-square error between the actual speed and the specified speed during the execution of tasks. Figure 17 shows the speed deviations during normal driving and when using each user interface. The results of a repeated measures one-way ANOVA showed that there were no significant differences among the no-task baseline, the tactile interface, and the visual and gestural interface ($F_{2,62} = 1.67$, $P > 0.05$). In addition, the tactile and new visual and gestural interfaces did not differ significantly ($F_{1,31} = 3.38$, $P > 0.05$).

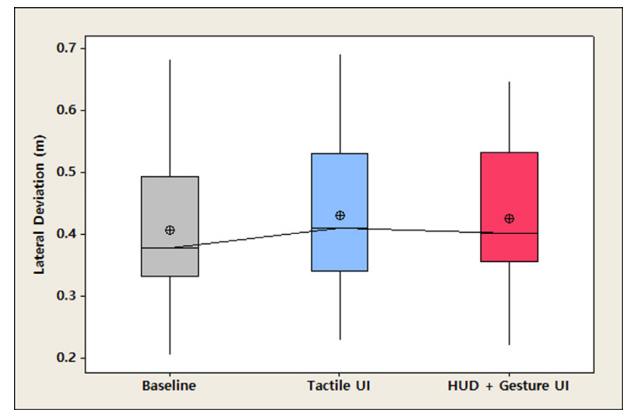

**Figure 19:** Lateral deviation effects on the three roads.

Figure 18 shows the speed deviations on the first easy road and the second difficult road. According to the results of a repeated measures one-way ANOVA for these two roads, there were no significant differences among the no-task baseline, tactile interface, and visual and gestural interface ($F_{2,62} = 0.47$, $P > 0.05$). In addition, there were no significant differences between the easy and difficult roads ($F_{1,31} = 2.67$, $P > 0.05$), operation and no operation ($F_{1,31} = 0.53$, $P > 0.05$), the baseline and the tactile interface ($F_{1,31} = 0.76$, $P > 0.05$), the baseline and the visual and gestural interface ($F_{1,31} = 0.10$, $P > 0.05$), or the tactile and visual and gestural interfaces ($F_{1,31} = 0.41$, $P > 0.05$).

### 5.2. Lateral deviation effect

Lateral deviation measures the driver's ability to maintain a central lane position and is a common measurement in other driver distraction studies. We computed lateral deviation as the root-mean-squared error of the lateral distance between the vehicle's centre and the lane centre. Figure 19 shows the results of an ANOVA indicating that the lateral deviation effect shows no significant changes among the three conditions ($F_{2,62} = 0.72$, $P > 0.05$). In addition, the tactile and new visual and gestural interfaces did not differ significantly ($F_{1,31} = 0.10$, $P > 0.05$).

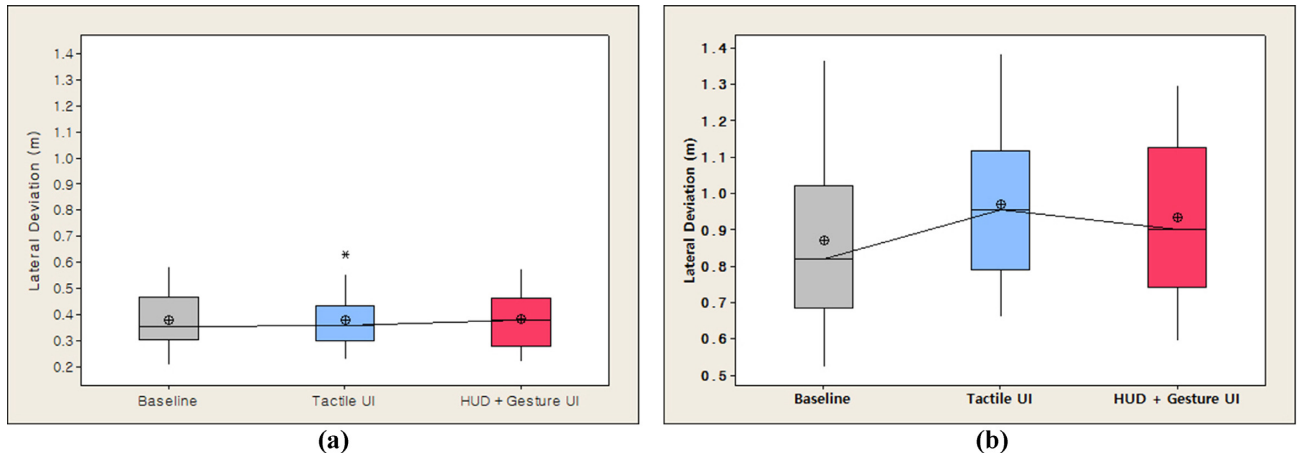

**Figure 20:** Lateral deviation effects: (a) on the first easy road and (b) on the second challenging road.

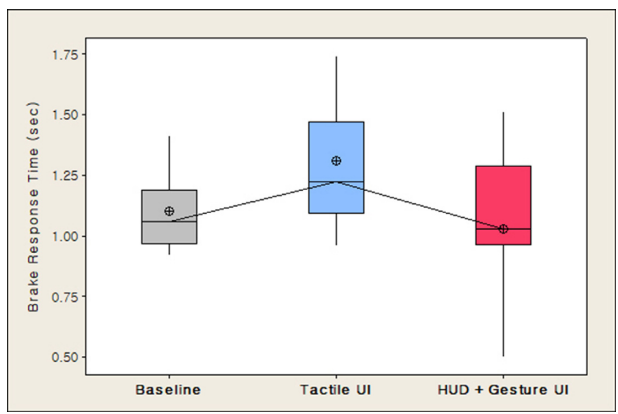

**Figure 21:** Brake response times on the three roads.

Figure 20 shows lateral deviations on the first easy road and second challenging road. According to the results of a repeated measures one-way ANOVA for the two roads, there were no significant differences among the no-task baseline, tactile interface, and visual and gestural interface ($F_{2,62} = 0.74, P > 0.05$). In addition, there were no significant differences between operation and no operation ($F_{1,31} = 1.40, P > 0.05$), the baseline and the tactile interface ($F_{1,31} = 1.27, P > 0.05$), the baseline and the visual and gestural interface ($F_{1,31} = 0.79, P > 0.05$), and the tactile and the visual and gestural interfaces ($F_{1,31} = 0.09, P > 0.05$). However, lateral deviations on the easy and difficult roads differed significantly ($F_{1,31} = 31.38, P < 0.001$).

### 5.3. Brake response time

Among the 31 tasks shown in Tables 1 and 2, 12 were conducted while driving under hazardous road condition with accidents, sharp corners, road construction, cut-ins, and tunnels. Figure 21 shows the mean brake response times for the baseline and two different interfaces. Unlike speed and lateral deviations, there was a significant overall effect of the user interface on brake response time ($F_{2,62} = 13.02, P < 0.001$). The differences between the tactile interface and the visual and gestural interface ($F_{1,31} = 21.08, P < 0.001$) and the baseline and the tactile interface ($F_{1,31} = 20.17, P < 0.001$) were significant. The response time for the tactile interface was 19% longer than that of the baseline; however, the difference between the baseline and the proposed visual and gestural interface was not significant ($F_{1,31} = 1.42, P > 0.05$). In addition, the response time for the visual and gestural interface was found to be almost the same as that of the baseline, indicating that a driver operating the device using our proposed interface can respond to emergencies as quickly as when driving a car without operating the device.

Figure 22 shows brake response times on the first easy road and the second challenging road. According to the results of a repeated measures one-way ANOVA for the two roads, there were no significant differences between the easy and difficult roads ($F_{1,31} = 0.90, P > 0.05$), operation and no operation ($F_{1,31} = 1.47, P > 0.05$), and the baseline and the visual and gestural interface ($F_{1,31} = 1.81, P > 0.05$). However, there were significant differences among the no-task baseline, tactile interface, and visual and gestural interface ($F_{2,62} = 14.30, P < 0.001$), between the baseline and the tactile interface ($F_{1,31} = 19.17, P < 0.001$), and between the tactile interface and the visual and gestural interface ($F_{1,31} = 24.58, P < 0.001$).

### 5.4. Task completion time

Task completion time was measured from the moment the verbal command was finished until 'end' was reported by the subject. In the tactile interface, the subject reported 'end' at the moment when their right hand returned to the steering wheel. There was found to be a significant effect for the interface type ($F_{1,31} = 32.68, P < 0.001$), such that the mean task completion time using visual and gestural interface was 19% faster than that using the tactile interface, as shown in Fig. 23.

Figure 24 shows task completion times on the first easy road and the second challenging road. According to the results of a repeated measures one-way ANOVA for the two roads, there were no significant differences between the easy and difficult roads ($F_{1,31} = 0.44, P > 0.05$). However, the tactile and the visual and gestural interfaces differed significantly ($F_{1,31} = 62.55, P < 0.001$).

### 5.5. Eye movements

Video sequences storing gaze information collected by the Ergoneers Dikablis Professional™ eye tracking and data acquisition system were analysed using D-Lab 3.0™ data acquisition and analysis software. Several areas of interest (AOIs) can be identified in the image shown to the driver. For a selected time interval, D-Lab analyses data to calculate the mean glance duration(s), which reflect the average glance duration in the direction of an AOI, the glance rate (s$^{-1}$), which is the glance frequency in an AOI, the AOI attention ratio (%), which is the percentage of glances at the AOI, mean fixation duration (ms) representing the length of time that a glance is fixed on an AOI, and horizontal and vertical eye activities (pixel) corresponding to standard deviations of the pupil in the X- and Y-axis in pixels. Figure 25 shows an example of eye tracking results. Although the driver gazes at the front field of view, it cannot be unambiguously said that the are attentive to the road. Since the menu is displayed on a HUD, it is natural that our interface shows a higher AOI attention ratio, higher mean fixation duration, and lower eye activities than when using the conventional tactile interface. Here, we present and compare the AOI attention ratios and vertical eye activities of the baseline and two interfaces in order to evaluate eye movements for later interface design. Note that the data on eye movements were collected for only 16 participants for roads 1 and 2.

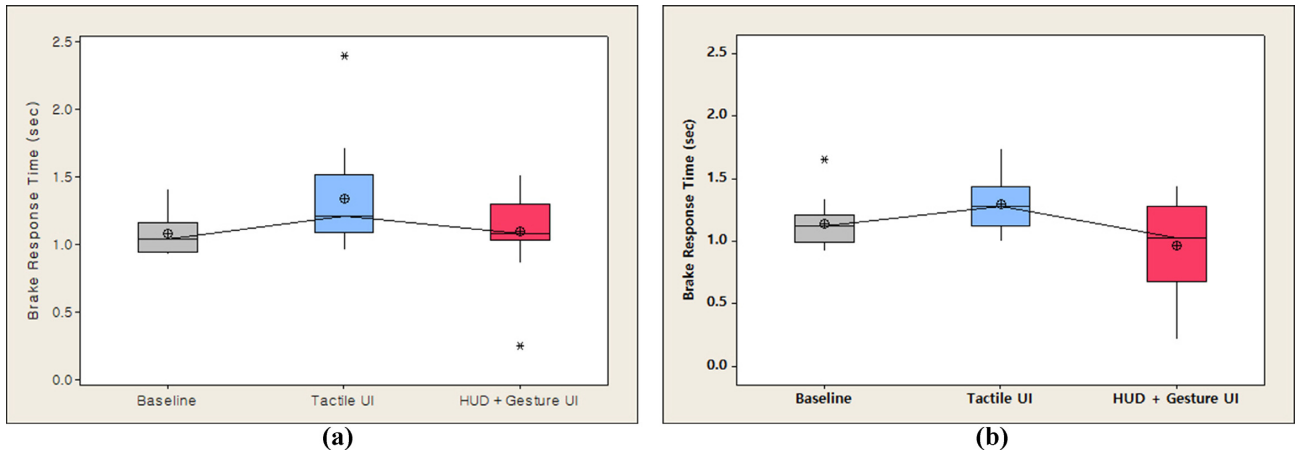

**Figure 22:** Brake response times: (a) on the first easy road and (b) on the second challenging road.

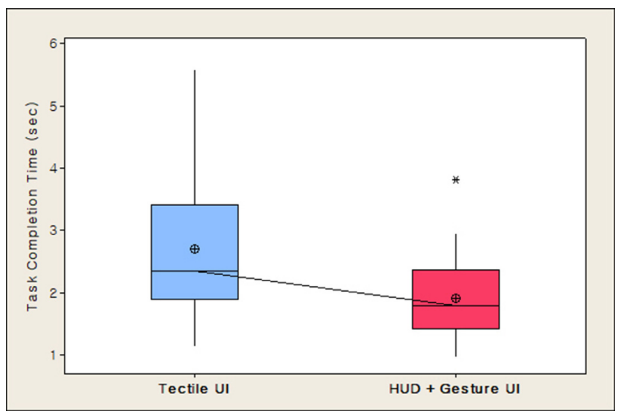

**Figure 23:** Secondary task completion times.

- AOI Attention Ratio: As shown in Fig. 26a, the attention ratio for the forward field of view was 86.06% (SD = 7.71) when driving the baseline, 79.72% (SD = 4.53) when using the tactile interface, and 86.89% (SD = 6.05) when using the visual and gestural interface. According to the results of a repeated measures one-way ANOVA, a significant main effect was detected for interface type ($F_{2,30} = 18.00$, $P < 0.001$). The visual and gestural interface showed a dwell time 8.25% higher than the tactile interface ($F_{1,15} = 26.77$, $P < 0.001$). Although differences between the baseline and tactile interface were significant ($F_{1,15} = 26.77$, $P < 0.001$), the differences between the baseline and the visual and gestural interface were not ($F_{1,15} = 0.65$, $P = 0.432 > 0.05$).
- Mean Fixation Duration: Figure 26b shows that the mean fixation duration was 238 ms (SD = 46.9) when driving the baseline, 232 ms (SD = 32.3) when using the tactile interface, and 232 ms (SD = 38.0) when using the visual and gestural interface. Interface type was therefore shown to have a significant effect ($F_{2,30} = 6.41$, $P < 0.01$). The differences between the tactile interface and the visual and gestural interface ($F_{1,15} = 9.32$, $P < 0.01$) and the baseline and the tactile interface ($F_{1,15} = 13.19$, $P < 0.01$) were significant. The mean fixation duration for the visual and gestural interface was 8% longer than that of the tactile interface. The difference between the baseline and the proposed visual and gestural interface was not significant ($F_{1,15} = 0.60$, $P > 0.05$).
- Horizontal Eye Activity: As shown in Fig. 26c, there was a significant overall effect of user interface on horizontal eye activity ($F_{2,30} = 7.61$, $P < 0.01$). The difference between driving without tasks and driving with tasks was significant ($F_{1,15} = 10.57$, $P < 0.01$). However, the difference between the tactile interface and the proposed visual and gestural interface was not significant ($F_{1,15} = 2.88$, $P > 0.05$).
- Vertical Eye Activity: Figure 26d shows that there was also a significant overall effect of user interface on vertical eye activity ($F_{2,30} = 85.59$, $P < 0.001$). Differences between the tactile interfaces and the visual and gestural interface ($F_{1,15} = 35.09$, $P < 0.001$) and between the tactile interfaces and the baseline ($F_{1,15} = 142.8$, $P < 0.001$) were significant. The difference between the tactile interface and the proposed visual and gestural interface was not significant ($F_{1,15} = 2.88$, $P > 0.05$).

### 5.6. Head movements

Figure 27 shows variations in head yaw, pitch, and roll for the subjects during task execution. The black line indicates the head movements when using the tactile interface, and the red line indicates the head movements with the new visual and gestural interface. The new interface involved almost no head movement, whereas the tactile interface yielded significant head movement. Figure 28 shows a box plot of the maximum values for head yaw, pitch, and roll during task execution. The mean head yaw when using the tactile interface was − 5.1° (SD = 4.6), whereas that with the new interface was 2.6° (SD = 3.0), demonstrating that head yaw was more to the right with the tactile interface ($F_{1,31} = 30.25$, $P < 0.05$). The mean head pitch when using the tactile interface was −4.6° (SD = 4.3), whereas that with the new interface was −1.4° (SD = 3.1), demonstrating that the head pitched down more with the tactile interface than with the new interface ($F_{1,31} = 9.92$, $P < 0.01$). The difference in head rolls between the baseline and the visual and gestural interface was not significant ($F_{1,31} = 2.43$, $P > 0.05$).

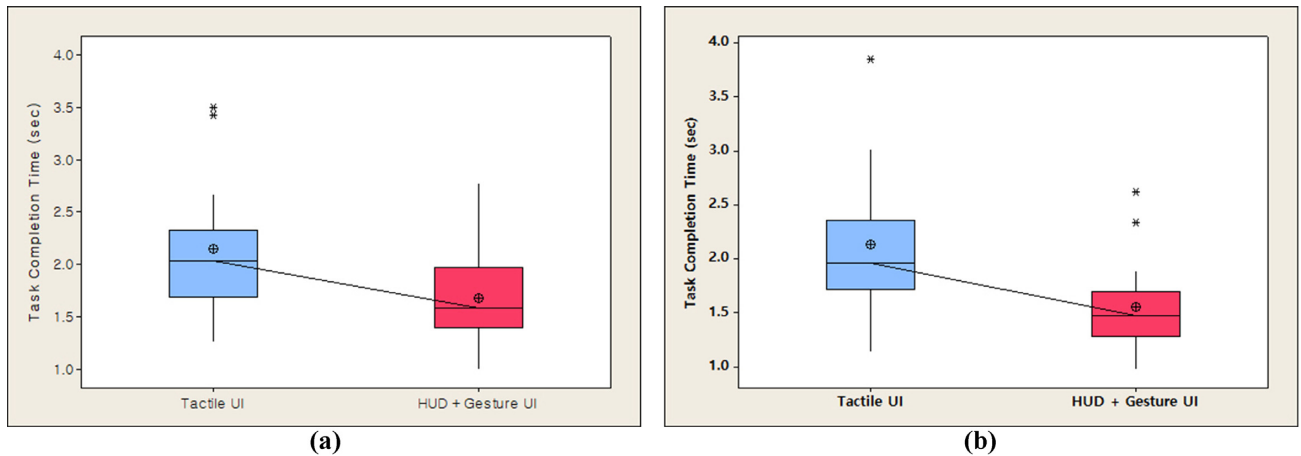

**Figure 24:** Secondary task completion times: (a) on the first easy road and (b) on the second challenging road.

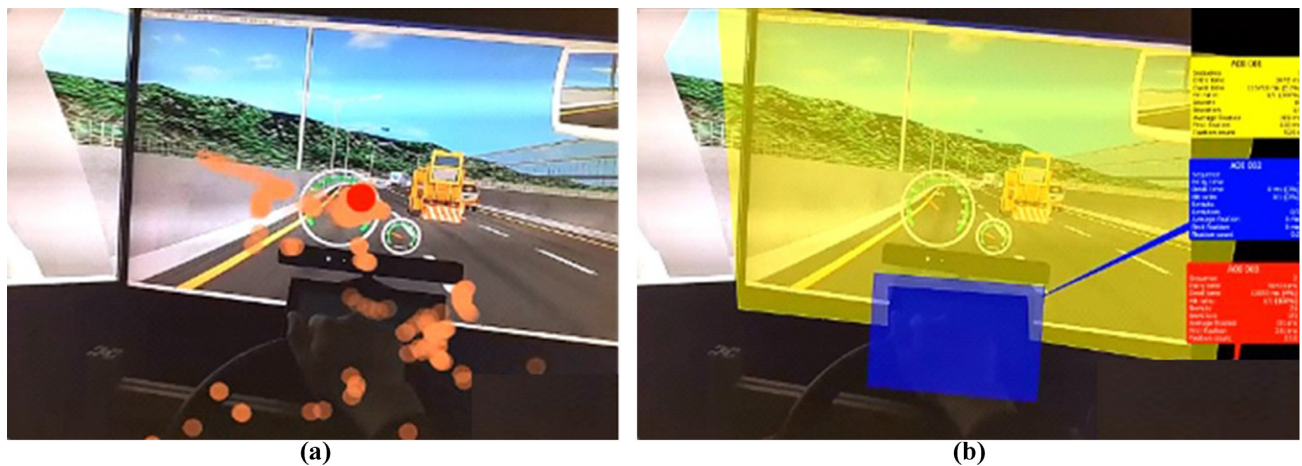

**Figure 25:** Eye tracking results: (a) eye fixation and (b) areas of interest.

### 5.7. Subjective evaluation

To determine the effect of task execution while driving, we assessed the level of difficulty for dependent variables, i.e. maintaining speed and lane, secondary task performance, setting confirmation, and looking forward while operating the device. The level of difficulty experienced by the drivers for each variable was evaluated based on a five-point Likert scale, where 1 = very difficult, 2 = difficult, 3 = neutral, 4 = easy, and 5 = very easy. A Wilcoxon signed-rank test was used for subjective evaluation through a questionnaire. As shown in Fig. 29, the test results indicated that the visual and gestural interface was more desirable than the tactile interface. In addition, Wilcoxon signed-rank tests determined that there was a statistically significant media increase in maintaining speed ($W = 0.0$, $P < 0.001$), maintaining lane ($W = 6.0$, $P < 0.001$), setting confirmation ($W = 0.0$, $P < 0.001$), task performance ($W = 17.0$, $P < 0.001$), and looking forward ($W = 0.0$, $P < 0.001$) when the visual and gestural interface was used.

## 6. Discussion

### 6.1. Overall experiment results

Visual distraction usually results in the driver taking their eyes off the road. HUD interfaces are designed to keep the eyes of the driver focused on the road ahead; however, visual distraction may still occur when users visually attend to the displayed information and miss cues important to the driving task. As traditional eyes-off-road measures are likely not well-suited for distraction due to HUDs, distraction needs to be measured in other ways. According to (Pampel & Gabbard, 2017), situation awareness, driver performance and other subjective measures can be affected by visual distraction; thus, they can be used as alternatives to traditional visual measures. One way to measure situation awareness is to probe whether the driver has recognized hazards, such as the lead car braking or an accident ahead. We measured the driver's braking response time for sudden dangers. Regarding driver performance, visual distraction can affect the driver's ability to keep a constant lane, speed, and gap and can affect lateral stability. In this study, we measured the maintenance of the requested speed, lateral deviation and task completion time. We also collected and analysed the subjective evaluation results.

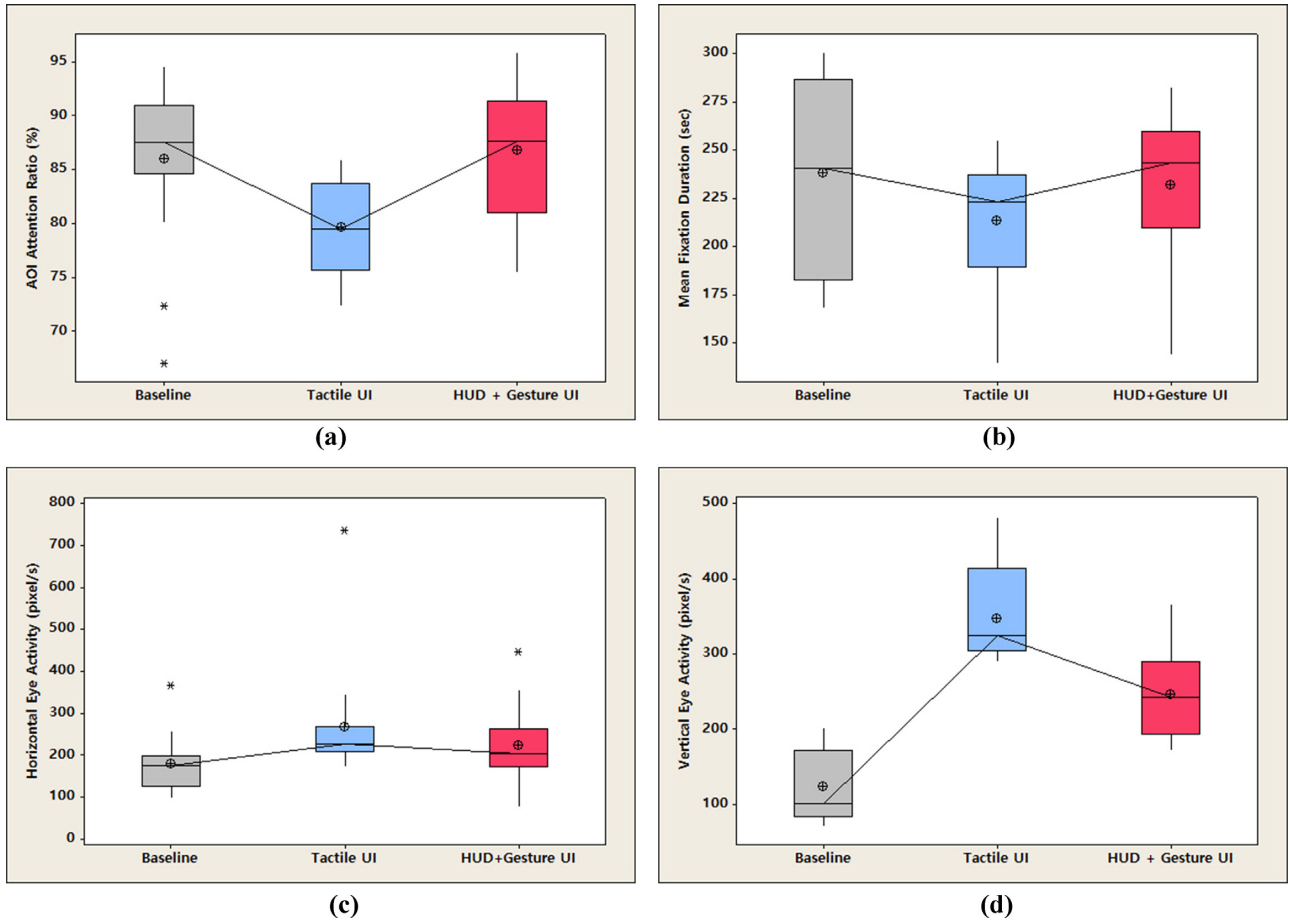

**Figure 26:** Eye movement measurement for the forward field of view: (a) AOI attention ratio, (b) mean fixation duration, (c) horizontal eye activity, and (d) vertical eye activity.

The experimental results for all roads, under normal and hazardous conditions, are presented for the traditional interface using tactile control and a HDD, and for the newly proposed interface using on-wheel gesture control and a HUD.

- For the performance of the primary task, i.e. driving, the proposed interface yielded a 3% higher speed-keeping rate and a 2% higher lane-keeping rate than the traditional interface. We found no significant difference between the two different interfaces; however, our interface yields a 20% faster brake response time.
- For the performance of secondary tasks, the proposed interface achieved a 34% faster task completion time than the traditional interface. These effects were more remarkable under hazardous road conditions, which were characterized by sharp curves and emergencies, than under normal road conditions.
- The traditional interface requires significant head movements both downwards and to the right in order to press the device control keys on the central console, whereas the proposed interface requires only finger gestures to control each device, allowing the drivers to keep their hands on the steering wheel. The display output can also be viewed on the HUD, requiring little head movement. Regarding eye movement, the new interface obtained an 8% longer dwell time for the forward field of view and an 8% higher average fixation count; however, we found no significant difference in the average fixation time.
- Based on the questionnaire, the newly proposed interface obtained a 31% better speed-keeping capability, 64% better lane-keeping capability, 64% better secondary task performance, 57% better setting confirmation capability, and 109% better forward-looking capability compared with the traditional interface.

The experimental results demonstrate that the proposed interface reduces visual and biomechanical distractions when locating and controlling devices compared with the traditional interface. This is considered to be beneficial for safe driving.

### 6.2. Limitations and future work

#### 6.2.1. Effects of age and gender

There are several factors to consider when designing an in-vehicle computing system (Burnett, 2008). The two most-often addressed in studies are the driver's age and gender. First, with regard to age, young drivers may be particularly skilled in the use of computing technology relative to the overall population, but are also more prone to risk taking. Also, the lack of driving experience causes a limited ability to divide attention and prioritize information sources. Conversely, older drivers often suffer from visual impairments that can cause various problems when using the in-vehicle display. Studies have shown that older drivers may take 1.5 to 2 times longer to read information from in-vehicle displays than younger drivers. Therefore, the size, luminance, contrast, and functional complexity of the information presented on the in-vehicle display are particularly important design factors. Second, gender-related effects on road safety attitudes and driver behaviour have been found in previous studies (Cordellieri et al., 2016). In general, men have higher crash rates than women. This gender difference is most pronounced between the ages of 20–29 years, after which it declines rapidly with age (SIRC, 2004).

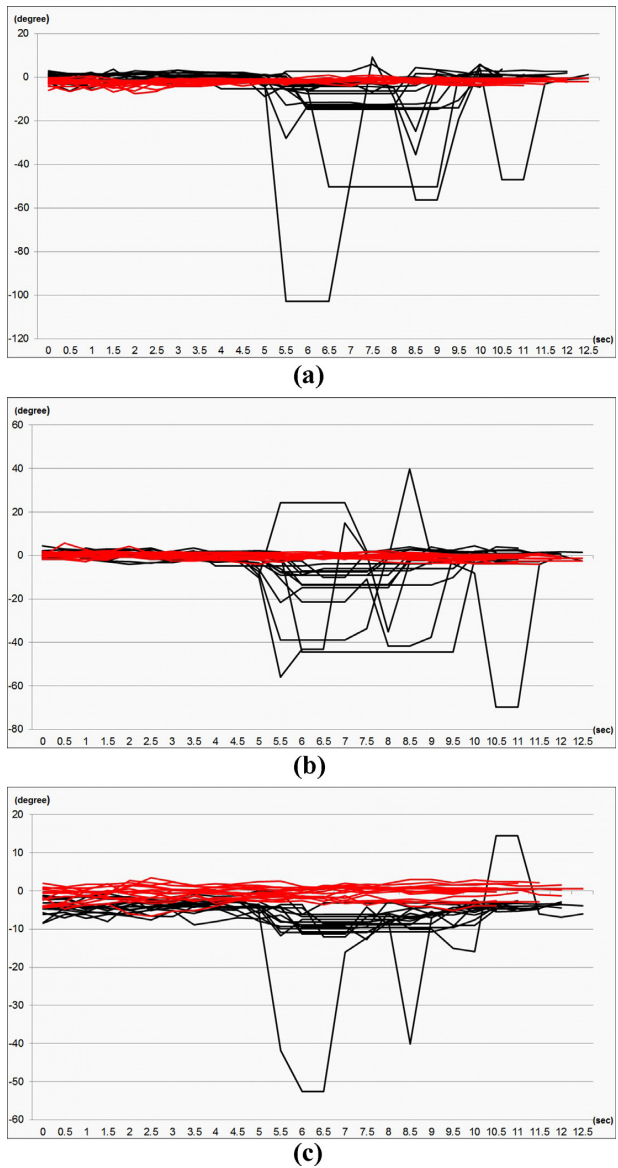

**Figure 27:** Head movements in all tasks (black lines for the tactile interface and red lines for the new visual and gestural interface): (a) yaw; (b) pitch; and (c) roll.

According to statistics provided by the Korean National Police Agency (2019), 32 million people held a driving licence in 2018, comprising 18 million men (58.2%) and 13 million women (41.8%). The average age of Korean drivers is 40.2 (SD = 12.2) years. However, in this study, 28 men and four women participated in the experiment. Their average age was 24.3 years (SD = 2.1), and their average driving experience was 3.4 months (SD = 1.3). Therefore, the average age and gender ratio of the subjects participating in the experiment is dissimilar to the average of all Korean drivers. The trend of the experimental results is expected to be the same in respect to age and gender, but this has not been confirmed, and the deviations are thus not known. This is one of the limitations of our study, and investigating the effects of age and gender for the population should form the subject of future work.

### 6.2.2. Effects of driving simulator

The driving simulator provides a controllable, safe, and cost-effective environment for data collection, making it a useful tool for studying driving behaviour (Risto & Martens, 2014). In a virtual environment, new forms of driver assistance and conduct experiments can be quickly implemented and tested in a controlled environment without needing to comply with road safety regulations. However, the results obtained from simulator research need to be evaluated with regard to their generalizability to the real world. Driver behaviour data collected in artificial scenarios under controlled conditions may not necessarily be similar to driver behaviour in real-world situations. It is therefore necessary to verify the validity of the simulator results.

Several means of validating driving simulator performance have been identified. Behavioural validity, which indicates how the driver's behaviour changes due to experimental conditions in a simulator and how this resembles changes in real-world driving, is very important (Knapper, Christoph, Hagenzieker, & Brookhuis, 2015). This is also distinguished by absolute and relative validity. Absolute validity is obtained when the numerical values measured in the simulator and the comparing methods are equivalent. Relative validity refers to values changing in the same direction and with comparable amplitude across methods. Concerning the usefulness of applying a driving simulator as a method for investigating driving, relative validity may, when carefully applied, suffice for generalizing to real-world driving.

In this work, we conducted experiments on a driving simulator using the proposed interface in various scenarios. However, in order to maintain a controlled environment and ensure the safety of participants, no experiments were conducted using real vehicles. It thus remains for future work to conduct experiments using a real vehicle and compare the results to verify the behavioural validity of the driving simulator.

### 6.2.3. Finger spreading gestures

Unlike other gestural interfaces, finger gestures have limitations in representing an object and/or an action for a task. The numbers or kinds of open fingers cannot be intuitively associated with objects and actions. A visual display is required to compensate for this limitation. In this study, a HUD was chosen over a screen on the central console. Unlike traditional tactile interfaces that require the driver to look at their fingertips while making a menu selection, the HUD helps to respond quickly by reducing gaze movement in an emergency. The proposed new finger gestures make it easier to handle emergencies when compared to normal space gestures, where the hands are not detached from the steering wheel.

We should note that the finger spreading method is basically a number input method, requiring a display to show the menus to be selected. A new type of interface that is less dependent on a display, and that does not require the driver to take their hands off the steering wheel, is evidently required.

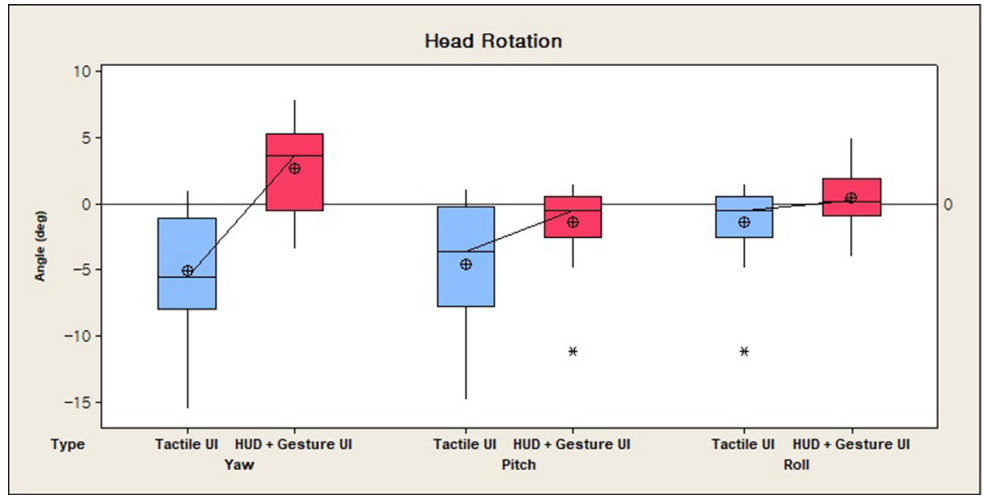

**Figure 28:** Head movement measurement: yaw, pitch, and roll angles.

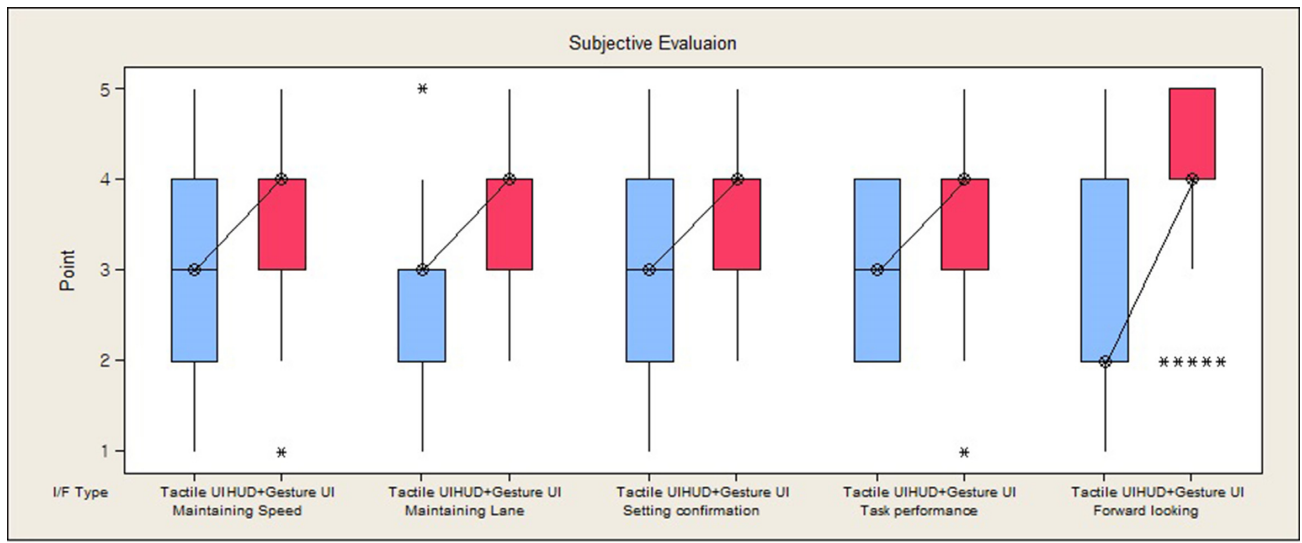

**Figure 29:** Level of difficulty in using two different interfaces based on the results of the questionnaire.

Moreover, in the present study, a P5 data glove was used for finger gesture recognition; however, this interface is only suitable for experiments and not for everyday use by drivers. Thus, a system needs to be developed that can recognize finger gestures when data gloves are not worn, the most promising method of which is computer vision technology using gesture recognition, in which robust algorithms have been developed to achieve a high recognition rate even under varying degrees of illumination. Finally, because newer vehicles often have additional button controls on the steering wheel, which are used as hotkeys to the central console controls, driver performance in using on-wheel buttons should be investigated and compared with the proposed on-wheel finger spreading gesture method.

## 7. Conclusion

In this study, we developed a user interface based around on-wheel gesture control and a HUD, which is expected to minimize visual and biomechanical distractions while controlling the audio and A/C systems on a central console when driving. Based on objective measurements and subjective evaluations of an experimental simulation, we compared the proposed system both quantitatively and qualitatively against a traditional interface utilizing tactile control and a HDD. Task completion rate, the time required for device control, the drivers' ability to maintain their speed and lane, and head and eye movements related to driver distraction were determined through objective measurements. Next, a questionnaire survey of the 15 subjects used was conducted to obtain subjective evaluations of the two types of interfaces. The levels of difficulty in terms of driving, performing secondary tasks, and looking forward while executing a task when driving were evaluated for both interfaces using the questionnaire. Our results show that the proposed interface reduces visual and biomechanical distractions for drivers when compared with the traditional interface.

## Acknowledgement

This research was supported by Basic Science Research Program through the National Research Foundation of Korea (NRF) funded by the Ministry of Education (NRF-2017R1D1A1B03036384).

## Conflict of interest statement

None declared.

## References

Angelini, L., Carrino, F., Carrino, S., Caon, M., Khaled, O. A., Baumgartner, J., Sonderegger, A., Lalanne, D., & Mugellini, E. (2014). Gesturing on the steering wheel: A user-elicited taxonomy. In *Proceedings of the 6th International Conference on Automotive User Interfaces and Interactive Vehicular Applications*, Seattle, WA, 17–19 September 2014, (pp. 1–8).

Angelini, L., Carrino, F., Carrino, S., Caon, M., Lalanne, D., Khaled, O. A., & Mugellini, E. (2013). Opportunistic synergy: A classifier fusion engine for micro-gesture recognition. In *Proceedings of the 5th International Conference on Automotive User Interfaces and Interactive Vehicular Applications*, Eindhoven, Netherlands, 28–30 October 2013, (pp. 30–37).

Bach, K. M., Jaeger, M. G., Skov, M. B., & Thomassen, N. G. (2008). You can touch, but you can't look: Interacting with in-vehicle systems. In *Proceedings of the SIGCHI Conference on Human Factors in Computing Systems*, Florence, Italy, April 2008, (pp. 1139–1148).

Bach, K. M., Jæger, M. G., Skov, M. B., & Thomassen, N. G. (2009). Interacting with in-vehicle systems: Understanding, measuring, and evaluating attention. In *Proceedings of the 23rd British HCI Group Annual Conference on People and Computers: Celebrating People and Technology*, Cambridge, UK, 1–5 September 2009, (pp. 453–462).

Barón, A., & Green, P. (2006). Safety and usability of speech interfaces for in-vehicle tasks while driving: A brief literature review. *Technical Report: University of Michigan Transportation Research Institute (UMTRI)*, February 2006.

Bellotti, F., De Gloria, A., Montanari, R., Dosio, N., & Morreale, D. (2005). COMUNICAR: Designing a multimedia, context-aware human–machine interface for cars. *Cognition, Technology & Work*, 7, 36–45.

Burnett, G. (2008). Designing and evaluating in-car user interfaces. In Zaphris, & C.S. Ang (ed.) *Human computer interaction: Concepts, methodologies, tools, and applications*, (pp. 532–551), Hershey, PA: Information Science Reference.

Choi, N. C., & Lee, S. H. (2015). Discomfort evaluation of truck ingress/egress motions based on biomechanical analysis. *Sensors*, 15(6), 13568–13590.

Cordellieri, P., Baralla, F., Ferlazzo, F., Sgalla, R., Piccardi, L., & Giannini, A. M. (2016). Gender effects in young road users on road safety attitudes, behaviors and risk perception. *Frontiers in Psychology*, 7, 1412.

Döring, T., Kern, D., Marshall, P., Pfeiffer, M., Schoning, J., Gruhn, V., & Schmidt, A. (2011). Gestural interaction on the steering wheel: Reducing the visual demand. In *Proceedings of the SIGCHI Conference on Human Factors in Computing Systems*, Vancouver, Canada, 7–12 May 2011, (pp. 483–492).

Endres, C., Schwartz, T., & Müller, C. A. (2011). Geremin: 2D microgestures for drivers based on electric field sensing. In *Proceedings of the 16th International Conference on Intelligent User Interfaces*, Palo Alto, CA, 13–16 February 2011, (pp. 327–330).

Eom, H., & Lee, S. H. (2015). Human-automation interaction design for adaptive cruise control systems of ground vehicles. *Sensors*, 15(6), 13916–13944.

Ergoneer Dikablis Glasses. (2016). Available online at http://www.ergoneers.com/en/hardware/eye-tracking/#dikablis, Accessed 30 May 2020.

Fukuda, T., Yokoi, K., Yabuki, N., & Motamedi, A. (2019). An indoor thermal environment design system for renovation using augmented reality. *Journal of Computational Design and Engineering*, 6(2), 179–188.

Gellatly, A. W. (1997). *The use of speech recognition technology in automotive applications (PhD thesis)*. Virginia Polytechnic Institute and State University, Blacksburg, VA.

González, I. E., Wobbrock, J. O., Chau, D. H., Faulring, A., & Myers, B. A. (2007). Eyes on the road, hands on the wheel: Thumb-based interaction techniques for input on steering wheels. In *Proceedings of the Graphics Interface*, Montreal, Canada, 28–30 May 2007, (pp. 95–102).

Ha, H., & Ko, K. (2015). A method for image-based shadow interaction with virtual objects, *Journal of Computational Design and Engineering*, 2, 26–37.

Hong, D., & Woo, W. (2008). Recent research trend of gesture-based user interfaces. *Telecommunications Review*, 18, 403–413.

Kang, M. S. (2012). *Gesture Interaction Design for Cars (M.S. thesis)*. Graduate School of Techno Design, Kookmin University.

Kiefer, R. J. (1998). Defining the "hud benefit time window". *Vision in Vehicles*, 6, 133–142.

Klauer, S., Dingus, T., Neale, V., Sudweeks, J., & Ramsey, D. (2006). *The impact of driver inattention on near-crash/crash risk: An analysis using the 100-car naturalistic driving study data*. Washington, DC, USA, (No. HS-810 594): National Highway Transportation Safety Administration.

Knapper, A., Christoph, M., Hagenzieker, M., & Brookhuis, K. (2015). Comparing a driving simulator to the real road regarding distracted driving speed. *European Journal of Transport and Infrastructure Research*, 15(2), 205–225.

Korean National Police Agency. (2019). Driver's license status. Available online at https://www.data.go.kr/dataset/15029967/fileData.do, Accessed 30 May 2020.

Koyama, S., Sugiura, Y., Ogata, M., Withana, A., Uema, Y., Honda, M, Yoshizu, S., Sannomiya, C., Nawa, K., & Inami, M. (2014) Multi-touch steering wheel for in-car tertiary applications using infrared sensors. In *Proceedings of the 5th Augmented Human International Conference*, March 2014, Article No. 5.

Lauber, F., Follmann, A., & Butz, A. (2014). What you see is what you touch: Visualizing touch screen interaction in the head-up display. In *Proceedings of the 2014 Conference on Designing Interactive Systems* (pp. 171–180).

Lee, S. H., & Ahn, D. R. (2015). Design and verification of driver interfaces for adaptive cruise control systems. *Journal of Mechanical Science and Technology*, 29(6), 2451–2460.

Lee, H., Jung, M., Lee, K. K., & Lee, S. H. (2017). A 3D human–machine integrated design and analysis framework for squat exercises with a Smith machine. *Sensors*, 17(2), 299.

Liu, Y. C., & Wen, M. H. (2004). Comparison of head-up display (HUD) vs. head-down display (HDD): Driving performance of commercial vehicle operators in Taiwan. *International Journal of Human-Computer Studies*, 61(5), 679–697.

Lee, S. H., Yoon, S. O., & Shin, J. H. (2015). On-wheel finger gesture control for in-vehicle systems on central consoles. In *Adjunct Proceedings of the 7th International Conference on Automotive User Interfaces and Interactive Vehicular Applications*, Nottingham, UK, 2015.09.01-03, (pp. 94–99).

Llaneras, R. E. (2000). *NHTSA driver distraction internet forum: Summary and proceeding*, Washington, DC, USA (No. HS-809 204): National Highway Transportation Safety Administration.

Neale, V. L., Dingus, T. A., Klauer, S. G., Sudweeks, J., & Goodman, M. (2005). *An overview of the 100-car naturalistic study and findings*. Washington, DC, USA: National Highway Transportation Safety Administration, Paper 05–0400.

Nielsen, J. (1995). 10 usability heuristics for user interface design, *Nielsen Norman Group*, 1(1).

Noy, Y. I., Lemoine, T. L., Klachan, C., & Burns, P. C. (2004). Task interruptability and duration as measures of visual distraction. *Applied Ergonomics*, 35, 207–213.

Pampel, S. M., & Gabbard, J. L. (2017). Measures of visual distraction in augmented reality interfaces. In *Workshop on Augmented Reality for Intelligent Vehicles*, September 24, 2017, Oldenburg, Germany.

Pfleging, B., Schneegass, S., & Schmidt, A. (2012). Multimodal interaction in the car: Combining speech and gestures on the steering wheel. In *Proceedings of the 4th International Conference on Automotive User Interfaces and Interactive Vehicular Applications*, Portsmouth, NH, 17–19 October 2012, (pp. 155–162).

Prinzel, L. J., III, & Risser, M. (2004). Head-up displays and attention capture. *NASA Technical Report NASA/TM-2004-213000*.

Ranney, T. A., Garrott, W. R., & Goodman, M. J. (2001). NHTSA driver distraction research: Past, present, and future (No. 2001-06-0177). SAE Technical Paper.

Rempel, D., Camilleri, M. J., & Lee, D. L. (2004). The design of hand gestures for human–computer interaction: Lessons from sign language interpreters. *International Journal of Human-Computer Studies*, 72(10), 728–735.

Riener, A. (2012). Gestural interaction in vehicular applications. *Computer*, 45, 42–47.

Risto, M., & Martens, M. H. (2014). Driver headway choice: A comparison between driving simulator and real-road driving. *Transportation Research Part F: Traffic Psychology and Behaviour*, 25, 1–9.

Shim, J. S., & Lee, S. H. (2016). A study on tactile and gestural controls of driver interfaces for in-vehicle systems. *Korean Journal of Computational Design and Engineering*, 21(1), 42–50.

Shneiderman, B. (1998). Eight golden rules of interface design. In *Designing the user interface*, 3rd ed, USA: Addison Wesley.

SIRC. (2004) Sex differences in driving and insurance risk: An analysis of the social and psychological differences between men and women that are relevant to their driving behaviour. *Social Issues Research Centre*. Oxford.

Skrypchuk, L., Langdon, P., Sawyer, B. D., & Clarkson, P. J. (2020). Unconstrained design: Improving multitasking with in-vehicle information systems through enhanced situation awareness. *Theoretical Issues in Ergonomics Science*, 21(2), 183–219.

Sojourner, R. J., & Antin, J. F. (1990). The effects of a simulated head-up display speedometer on perceptual task performance. *Human Factors*, 32(3), 329–339.

Son, J., Jang, H., & Choi, Y. (2019). Tangible interface for shape modeling by block assembly of wirelessly connected blocks. *Journal of Computational Design and Engineering*, 6(4), 542–550.

Stutts, J. C., Feaganes, J., Rodgman, E., Hamlett, C., Meadows, T., Reinfurt, D., Gish, K., & Staplin, L. (2003). *Distractions in everyday driving*. Washington, DC, USA, (No. HS-043 573): AAA Foundation for Traffic Safety.

Stutts, J. C., Reinfurt, D. W., Staplin, L., & Rodgman, E. A. (2001). *The role of driver distraction in traffic crashes*. Washington, DC, USA: AAA Foundation for Traffic Safety.

Sun, C., Hu, W., & Xu, D. (2019). Navigation modes, operation methods, observation scales and background options in UI design for high learning performance in VR-based architectural applications. *Journal of Computational Design and Engineering*, 6(2), 189–196.

Takahashi, R., Suzuki, H., Chew, J.Y., Ohtake, Y., Nagai, Y., & Ohtomi, K. (2018). A system for three-dimensional gaze fixation analysis using eye tracking glasses. *Journal of Computational Design and Engineering*, 5(4), 449–457.

Tognazzini, B. (2014). First principles of interaction design (revised & expanded). AskTog. Available online at https://asktog.com/atc/principles-of-interaction-design/, Accessed 30 May 2020.

Tsimhoni, O., & Green, P. (2001). Visual demand of driving and the execution of display-intensive, in-vehicle tasks. In *Proceedings of the Human Factors and Ergonomics Society 45th Annual Meeting*, October 2001, (pp. 1586–1590).

UC-win/Road. (2019). Available online at https://www.forum8.co.jp/ENGLISH/uc-win/ucwin-road-e1.htm, Accessed 30 May 2020.

Ulrich, T. A., Spielman, Z., Holmberg, J., Hoover, C., Sanders, N., Gohil, K., & Werner, S. (2013). Playing charades with your car – The potential of free-form and contact-based gestural interfaces for human vehicle interaction. In *Proceedings of the Human Factors and Ergonomics Society Annual Meeting*, 1–3 October 2013, (Vol. 57, No. 1, pp. 1643–1647).

Wachs, J. P., Kölsch, M., Stern, H., & Edan, Y. (2011). Vision-based hand-gesture applications. *Communications of the ACM*, 54(2), 60–71.

Weinberg, G., Harsham, B., & Medenica, Z. (2011). Evaluating the usability of a head-up display for selection from choice lists in cars. In *Proceedings of the 3rd International Conference on Automotive User Interfaces and Interactive Vehicular Applications*(pp. 39–46).

Werner, S. (2014). The steering wheel as a touch interface: Using thumb-based gesture interfaces as control inputs while driving. In *Proceedings of the 6th International Conference on Automotive User Interfaces and Interactive Vehicular Applications*, 17–19 September 2014, (pp. 1–4), Seattle, WA.